\documentclass[review,sort&compress]{elsarticle}

\usepackage{fullpage}
\usepackage{amssymb}
\usepackage{graphicx}
\usepackage{subfig}
\usepackage{enumerate}
\usepackage{amsmath,mathtools}
\usepackage{amsfonts,bm}
\usepackage{todonotes}
\usepackage[commentmarkup=footnote]{changes}
\usepackage{stmaryrd}
\usepackage{siunitx}
\usepackage{caption}
\usepackage{array}
\usepackage{booktabs}
\usepackage{nomencl}
\usepackage{lineno}
\usepackage{url}
\usepackage{natbib}
\usepackage[strings]{underscore}
\newcommand{\region}[1]{\langle #1 \rangle}

\newcommand{\bbR}{\mathbb {R}}

\newcommand{\bu}{\mathbf{u}}
\newcommand{\bq}{\mathbf{q}}
\newcommand{\bc}{\bm{c}}

\newcommand{\cE}{\mathcal{E}}
\newcommand{\cO}{\mathcal{R}}
\newcommand{\cR}{\mathcal{Q}}
\newcommand{\cQ}{\mathcal{Q}}
\newcommand{\cG}{\mathcal{G}}
\newcommand{\cX}{\mathcal{X}}

\newcommand{\cD}{\mathcal{D}}
\newcommand{\cL}{\mathcal{L}}
\newcommand{\vnorm}[1]{|#1|}
\newcommand{\tnorm}[1]{\|#1\|}

\newcommand{\rstt}{\mathcal{R}}

\renewcommand\nomgroup[1]{%
  \item[\itshape%\bfseries
  \ifstrequal{#1}{A}{Symbols}{%
  \ifstrequal{#1}{B}{Roman Letters}{%
  \ifstrequal{#1}{C}{Greek Letters}{%
  \ifstrequal{#1}{D}{Abbreviations}{}}}}%
]}

\usepackage{xpatch}
\xpatchcmd{\thenomenclature}{\section*{\nomname}
}{}{\typeout{Success}}{\typeout{Failure}}
\makenomenclature
\RequirePackage{ifthen}
\printnomenclature[0.8cm]

\journal{Computer Methods in Applied Mechanics and Engineering}

\graphicspath{{./figs/}}

\begin{document}

\begin{frontmatter}

\title{An equivariant neural operator for developing nonlocal tensorial constitutive models}

\author[fi]{Jiequn Han\fnref{contribution}}%
\ead{jiequnhan@gmail.com}

\author[vt]{Xu-Hui Zhou\fnref{contribution}}%
\ead{xuhuizhou@vt.edu}

\author[vt]{Heng Xiao\corref{cor1}}%
\ead{hengxiao@vt.edu}

\fntext[contribution]{Contributed equally}
\cortext[cor1]{Corresponding author.}

\affiliation[fi]{organization={Center for Computational Mathematics, Flatiron Institute},
            city={New York},
            postcode={10010}, 
            state={NY},
            country={USA}}
\affiliation[vt]{organization={Kevin T. Crofton Department of Aerospace and Ocean Engineering, Virginia Tech},
            city={Blacksburg},
            postcode={24060}, 
            state={VA},
            country={USA}}

\begin{abstract}
Developing robust constitutive models is a fundamental and longstanding problem for accelerating the simulation of complicated physics. Machine learning provides promising tools to construct constitutive models based on various calibration data. In this work, we propose a neural operator to develop nonlocal constitutive models for tensorial quantities through a vector-cloud neural network with equivariance (VCNN-e). The VCNN-e respects all the invariance properties desired by constitutive models, faithfully reflects the region of influence in physics, and is applicable to different spatial resolutions. By design, the model guarantees that the predicted tensor is invariant to the frame translation and ordering (permutation) of the neighboring points. Furthermore, it is equivariant to the frame rotation, i.e., the output tensor co-rotates with the coordinate frame. We evaluate the VCNN-e by using it to emulate the Reynolds stress transport model for turbulent flows, which directly computes the Reynolds stress tensor to close the Reynolds-averaged Navier--Stokes (RANS) equations. The evaluation is performed in two situations: (1) emulating the Reynolds stress model through synthetic data generated from the Reynolds stress transport equations with closure models, and (2) predicting the Reynolds stress by learning from data generated from direct numerical simulations. Such a priori evaluations of the proposed network pave the way for developing and calibrating robust and nonlocal, non-equilibrium closure models for the RANS equations.
\end{abstract}

\begin{keyword}
neural operator \sep nonlocal closure model \sep constitutive modeling  \sep invariance and equivariance \sep deep learning
\end{keyword}

\end{frontmatter}

\section{Introduction}
\label{sec:intro}
The first principle-based simulations of many practical problems arising from scientific and engineering applications are prohibitively expensive due to their multiscale nature. The constitutive relationship builds an effective bridge between different scales to mitigate this difficulty by simplifying and approximating the unresolved process at the microscopic scale with the resolved process in order to accelerate the simulations at the macroscopic scale. 
Taking industrial computational fluid dynamics (CFD) for instance, simulating the Reynolds averaged Navier--Stokes (RANS) equations for the mean flows fields are of ultimate interest in many engineering design tasks. To this end, constitutive models such as eddy viscosity models~\cite{spalart92one,launder74application} and Reynolds stress models~\cite{launder1975progress} are often introduced to describe unresolved turbulence and close the RANS equations.
As such, they are also referred to as ``closure models'', a term that is used interchangeably with ``constitutive models'' in this paper. 
Other examples include developing nonlinear stress-strain constitutive models to describe nonlinear elasticity~\cite{coleman1967thermodynamics} and developing hydrodynamic models to approximate the motion of particles described by the kinetic equations or non-Newtonian fluids~\cite{han2019uniformly,lei2020machine}.
Despite the enormous growth of available computational resources in the past decades, even to this day, such closure models are still the backbone of practical engineering computations.

\subsection{Invariance properties of constitutive models}

In order to capture the essence of the physical phenomenon, the closure model should ideally express fundamental principles obeying physical constraints and meanwhile be as universally accurate as possible in the considered conditions. Traditional constitutive modeling usually relies on simple parametric models in combination with physical insight. However, as the number of involved state variables increases, it becomes more and more difficult to specify the form of the closure model and calibrate the parameters systematically, an essential obstacle known as the curse of dimensionality. 
The development of machine learning techniques in recent years, especially deep learning, has provided new tools to extract complicated relationships from massive data in an efficient and flexible way, thus a new possibility to data-driven constitutive modeling~\cite{han2021machine}. 
In light of such strengths and promises, many machine learning-based constitutive models have been developed in the past few years in a wide range of application areas such as turbulence modeling~\cite{ling16reynolds,wang17physics-informed,wu2018physics-informed,schmelzer2020discovery} and computational mechanics~\cite{bock2019review,huang2020learning,xu2021learning,masi2021thermodynamics,scoggins2021machine}.

In some cases of physical modeling, it has been observed that machine learning-based models tend to achieve the best performance when it respects exactly the physical constraints and symmetries ~\cite{han2021machine,wu2019representation,zafar2020convolutional,taghizadeh2020turbulence,wang2021incorporating}.
However, unlike the traditional constitutive modeling in which conventional wisdom can help restrict the set of admissible functions to obey physical constraints, such as frame independence~\cite{pope1975more,gatski1993on,speziale1991modelling}, it is not straightforward to restrict flexible machine learning models like neural networks that take the raw state variables as input to be always genuinely physical.
The challenge becomes more demanding when the constitutive relationship is nonlocal, i.e.,  the closure variable at location $\bm{x}_i$ depends on the values of the resolved variables in a neighborhood region of $\bm{x}_i$ rather than the value at $\bm{x}_i$ solely.
This situation is most evidently seen in turbulence models. The unresolved turbulent velocity fluctuation (which can be partly characterized by its second-order statistics, i.e., the Reynolds stress) depends on an upstream neighborhood due to the turbulent transport~\cite{gatski1996simulation}. In such circumstances, an ideal closure model should be indifferent to the choice of material frame and the representation of resolved variables in the nonlocal region. In other words, the constitutive relationship should remain invariant when the coordinate system is translated or the index order of the discretized variable in the neighborhood is permuted. Depending on the type of the closure variable, the constitutive relationship should also remain invariant or equivariant if the coordinate system is rotated. A recent work of Gao et al.~\cite{gao2020roteqnet} considers specifically the rotational invariance/equivariance of a local relationship between a pair of tensors in fluid dynamics. However, the technique developed therein is insufficient to handle the nonlocal relationship.
In a different vein, many recent works~\citep[see, e.g.][]{long2018pde,sun2020surrogate,kim19deep,lu2021learning,ma2020machine,ribeiro2020deepcfd,li2020neural,li2021fourier} used machine learning models to approximate global operators or surrogate models defined by the PDEs, which also hold the promise to nonlocal constitutive modeling. However, the objectivity of these modeling approaches, such as frame independence and permutational invariance mentioned above, has rarely been discussed.
How to ensure that the machine learning models obey various physical constraints in the context of nonlocal constitutive modeling remains largely an open question.
 
\subsection{Vector-cloud neural network with equivariance (VCNN-e)}
 
One line of efforts to impose all the aforementioned symmetries in data-driven modeling is Deep Potential~\cite{han2018deep,zhang2018end} for interatomic potential and the vector-cloud neural network for nonlocal constitutive modeling~\cite{zhou2022frame}. At a high level, the original model consists of two modules, an embedding network and a fitting network. The former extracts features that are translational, rotational, and permutational invariant from the raw input, and the latter predicts the output with the desired symmetry. 
This work focuses on generalizing the vector-cloud neural network (VCNN)~\cite{zhou2022frame} to develop a nonlocal closure model for \emph{tensorial} output quantities.
Compared to VCNN, in which the closure variable is a rotational invariant scalar~\cite{zhou2022frame}, the main innovation of the present work is to tackle the closure problem in which the closure variable is a rotational equivariant tensor. To this end, we adopt the idea of Sommers et al.~\cite{sommers2020raman} to equip the VCNN with equivariance for tensorial outputs while preserving its general approximation capability. Furthermore, we extend the VCNN equipped with equivariance to correctly represent the function space that is uniquely required for turbulence modeling, as illustrated in Fig.~\ref{fig:NN-architecture}(c). In so doing, we obtain a vector-cloud neural network with equivariance (VCNN-e) that satisfies all the symmetries and function space requirements for modeling Reynolds stresses. To demonstrate the merits of the proposed VCNN-e framework, we test it for (1) emulating a Reynolds stress transport equation via synthetic data generated by solving the equation with closure models, and (2) predicting the Reynolds stress through data from direct numerical simulations. The latter test differs from the former in that the Reynolds stress from DNS data is not described by an explicit tensor transport equation and is a more realistic test.
Through the validation on a family of parameterized periodic hill geometries, we showcase that the proposed vector-cloud neural network with equivariance can be trained flexibly from data and meanwhile capture the nonlocal physics of tensor transport accurately. 
Among other potential applications, the work paves the way for further development towards frame-independent, nonlocal, data-driven Reynolds stress closure models for the RANS equations.

\section{Problem Statement and Methodology}
\label{sec:method}

In this study, we intend to learn the nonlocal turbulence model for closing the Reynolds-averaged Navier--Stokes (RANS) equations. For industrial applications, the RANS models are still the workhorse as they do not resolve all scales and are much less expensive. The RANS equations are derived by performing Reynolds decomposition and ensemble averaging on the Navier--Stokes (N--S) equations, and are very similar to the original equations. However, the additional nonlinear terms in the momentum equations, namely Reynolds stress, are unknown and must be modeled. Here we propose using an equivariant neural operator to model the Reynolds stress term as a functional of the nonlocal mean flow fields instead of solving a transport PDE. We call it an operator because the proposed neural network model takes an discretized function as input and can adapt to different discretizations. The turbulence model in this work merely serves as an example of tensorial constitutive models described by transport PDEs, while the method is generally applicable for any nonlocal tensorial constitutive models.

For turbulent flows, the mean flow fields (velocity $\mathbf{u}$ and pressure $p$) are described by the following RANS equations:
\begin{equation}
    {\mathbf{u}} \cdot \nabla {\mathbf{u}} - \nu \nabla^2 \mathbf{u} = -\frac{1}{\rho} \nabla {p} + \nabla \cdot \rstt,
\end{equation}
where $\rho$ and $\nu$ is density and kinematic viscosity, respectively, and $\rstt$ is the Reynolds stress. To close the RANS equations, we model
the Reynolds stress $\rstt$ by a constitutive transport equation of the following form such that $\rstt$ can be determined from the mean flow fields:
\begin{equation}
    \label{eq:rstm}
    \mathbf{u} \cdot \nabla \rstt - \nabla \cdot \left[(\nu + \nu_t)\nabla \rstt \right] =  \mathcal{P} + \Phi - E,
\end{equation}
where $\nu_t$ denotes turbulent eddy viscosity, and $\mathcal{P}$, $\Phi$, $E$ denote production, pressure--strain-rate, and dissipation, respectively. The terms $\nu_t, \Phi, E$ need to be additionally modeled (see examples in Section~\ref{sec:results_synthetic}).

\begin{figure}[!htb]
\centering
\includegraphics[width=0.99\textwidth]{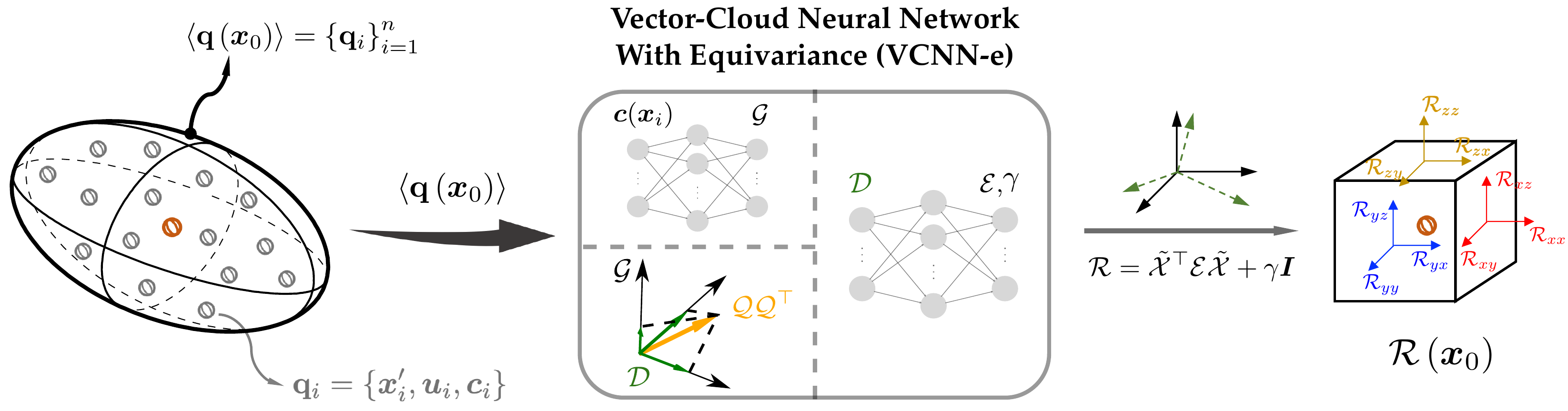}
  \caption{
  Schematic of the vector-cloud neural netowork with equivariance (VCNN-e) for nonlocal tensorial constitutive modeling, showing a region-to-point mapping from a cloud (ellipsoid) of feature vectors $\region{\mathbf{q}(\bm{x}_0)}$ to the tensorial closure quantity $\rstt(\bm{x}_0)$ at center of the ellipsoid. Details on the architecture of VCNN-e are shown subsequently in Fig.~\protect\ref{fig:NN-architecture}.
  }
  \label{fig:NN-schematic}
\end{figure}

Considering both the nonlocal physics embodied in the transport PDE~\eqref{eq:rstm} and feasibility for implementation in CFD solvers, the network should form a region-to-point mapping $\region{\mathbf{q}(\bm{x}_0)} \mapsto \rstt(\bm{x}_0)$ shown in Fig.~\ref{fig:NN-schematic}, where $\mathbf{q}$ is the feature vector (we always assume column vectors in this paper), and $\region{\mathbf{q}(\bm{x}_0)}$ indicates the collection of features $\{\bq_i\}_{i=1}^n$ on $n$ points sampled from the region around $\bm{x}_0$ (referred to as \emph{cloud}). The extent of the cloud is determined by the velocity $\bu$ at the point of interest according to the region of physical influence (see the detailed expression in~\cite{zhou2022frame}).
The number of points $n$ in a cloud can vary from location to location, which requires our model to adapt to different discretizations.
The feature vector $\mathbf{q}$ attached to each point is chosen to include the relative coordinate $\bm{x}' = \bm{x} - \bm{x}_0$, flow velocity $\bu$, and additional seven scalar quantities $\bc = [\theta, \mathsf{u}, s, b, \eta,  r, r']^\top$.
These scalar features are the same as those used in \cite{zhou2022frame}, and we briefly recall their definitions for completeness.
\begin{enumerate}[(1)]
    \item cell volume $\theta$, which represents the weight for each point in mesh-based methods;
    \item velocity magnitude~$\mathsf{u}=|\bm{u}|$ and strain rate magnitude~$s=\| \bm{S}\| = \tnorm{\nabla \mathbf{u} + (\nabla \mathbf{u})^\top}$, the latter of which often appears in various turbulence and transition models;
    \item boundary cell indicator $b$ and wall distance~$\eta$ after normalization by boundary layer thickness scale $\delta^*$ and capped by 1. These two variables help the closure model to distinguish between the mapping described by the differential PDE and wall model (boundary condition);
    \item proximity~$r$ (inverse of relative distance) to the center point of the cloud and proximity $\displaystyle r'$ defined based on the angle between $-\bu^\top$ and  $\bm{x}'$ accounting for the alignment between the convection and the relative position of the point in its cloud.
\end{enumerate}
According to the above definition, the feature vector $\bq$ is $3+3+7=13$ dimensional, and we define the input data matrix as $\cQ = [\mathbf{q}_1, \ldots, \mathbf{q}_n]^\top\in \mathbb{R}^{n\times 13}$. Note that all the input features are first non-dimensionalized by the characteristic scales and then fed into the neural network. As such the closure model can handle dynamically similar systems, e.g., training and predicting systems at similar Reynolds numbers despite with vastly different length and time scales. 

\begin{figure*}[!htb]
\centering
\includegraphics[width=0.99\textwidth]{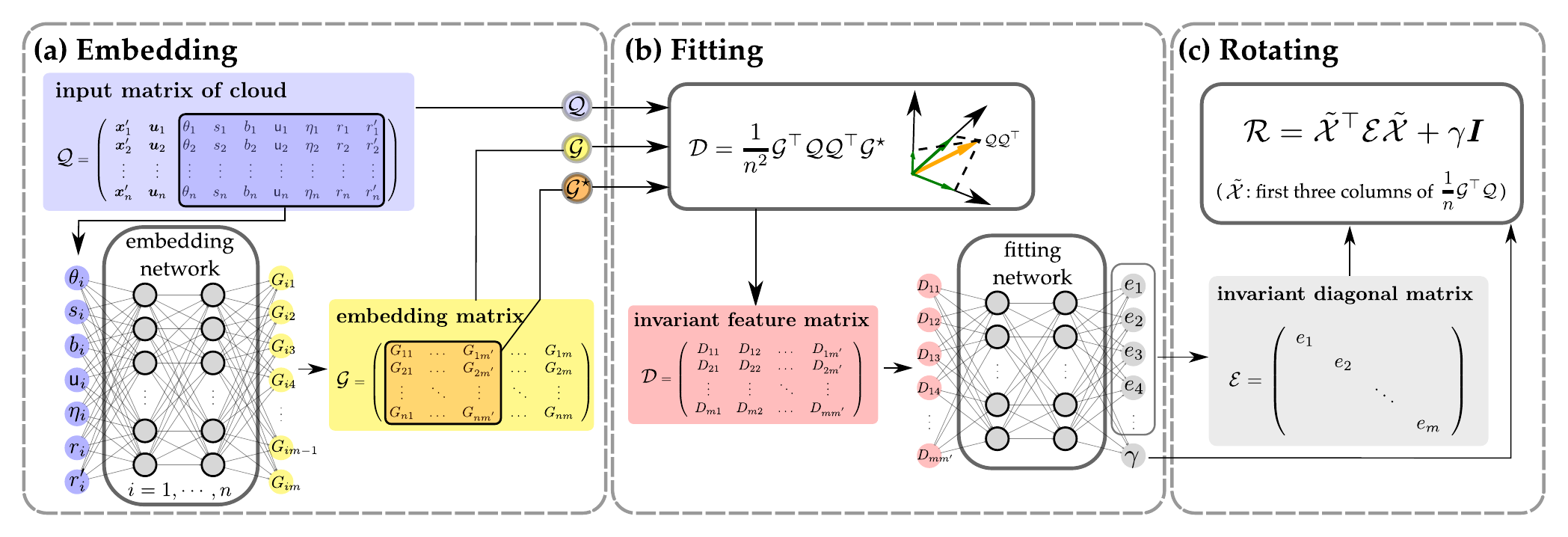}
  \caption{
  Detailed architecture of the vector-cloud neural network with equivariance (VCNN-e) for predicting a targeting tensor $\rstt$:
  (a) embed the frame-independent features $\{\bm{c}_i\}_{i=1}^n$ to form the embedding matrix $\mathcal{G} \in \mathbb{R}^{n \times m}$;
  (b) project the pairwise inner-product matrix $\mathcal{Q} \mathcal{Q}^\top$ to the learned embedding matrix $\mathcal{G}$ and its submatrix $\mathcal{G}^\star$ to yield an invariant feature matrix $\mathcal{D} \in \mathbb{R}^ {m \times m'}$;
  flatten and feed the feature matrix $\mathcal{D}$ into the fitting network to predict the diagonal matrix $\cE$ and scalar $\gamma$; 
  (c) rotate $\cE$ through the embedded coordinates $\tilde{X}$ and get the final prediction of $\rstt$.
  The constitutive mapping $\mathbf{u}(\bm{x}) \mapsto \rstt$ based on VCNN-e is invariant to both the frame translation and the ordering of points in the cloud, and equivariant to the frame rotation.
  The VCNN-e differs from its predecessor VCNN~\cite{zhou2022frame} in that the developed rotating panel (c) generalizes scalar output to tensor and ensures its equivariance.
  \label{fig:NN-architecture}
  }
\end{figure*}

The vector-cloud neural network with equivariance (VCNN-e) for modeling Reynolds stress tensor consists of three modules at a high level: an embedding module, a fitting module, and a rotating module.
The goal of the first module (Fig.~\ref{fig:NN-architecture}a) is to find a representation of the input data that is translational, rotational, and permutational invariant. 
To this end, we introduce a set of $m$ basis functions $\{\phi_k(\bc_i)\}_{k=1}^m$, implemented by an embedding neural network with $m$-dimensional output, for the scalar quantities $\bc_i$ (note the input $\bc_i$ is the frame-independent part of the feature vector $\bq_i$) and define $\cL_{k j} = \frac1n\sum_{i=1}^n \phi_k(\bc_i) \, \cR_{ij}$, $k=1, \ldots, m, j=1, \ldots, 13$. The summation over the point index $i$ removes the dependence of $\cL$ on the ordering and makes it permutational invariant. The normalization by $n$ allows a different number of sampled points in the cloud in the training or testing data and guarantees the resulting VCNN-e is resolution adaptive.
If we define a matrix $\cG_{ik} = \phi_k(\bc_i)$, the permutation invariant transformation above can be written as $\cL=\frac{1}{n} \cG^\top \cR$. 
Similarly, we define $\cL^\star = \frac{1}{n} \cG^{\star\top} \cR$, with $\cG^{\star}$ being the first $m'$ columns of $\cG$, and $\cL^*$ is also permutational invariant. 
Here we choose $\cG^{\star}$ as a subset of $\cG$ instead of $\cG$ itself mainly in order to save the computational cost in the next step without sacrificing the accuracy.
Next, we define $\cD = \cL \cL^{\star\top} = \frac{1}{n^2}\cG^\top\cR\cR^\top\cG^\star$, which is translational, rotational, and permutational invariant since both $\cL$ and $\cL^{\star\top}$ are.
We can view $\cD$ as a faithful extraction of information from the feature vectors in the cloud via projecting $\cR\cR^\top$ onto the learned basis $\cG$.
Then the second module (Fig.~\ref{fig:NN-architecture}b) takes $\cD$ as the input and fits an invariant diagonal matrix $\mathcal{E}$.
We construct a fitting network $f_\text{fit}: \cD \mapsto [e_1,\cdots,e_m, \gamma]^\top$ that maps $\cD$ to $(m+1)$-dimensional output, and define a diagonal matrix $\cE\in \bbR^{m\times m}$ taking $e_1,\cdots,e_m$ as the diagonal elements. 
Finally, in the third module (Fig.~\ref{fig:NN-architecture}c), we re-incorporate the embedded coordinates $\tilde{\cX}=\frac1n\cG^\top\cX\in\mathbb{R}^{m\times 3}$ from the first module where $\cX=[\bm{x}_1, \ldots, \bm{x}_n]^\top\in\mathbb{R}^{n\times 3}$ and rotate properly to get the final tensor prediction through
\begin{equation}
\label{eq:xex}
    \rstt = \tilde{\cX}^\top\cE\tilde{\cX} + \gamma \bm{I},
\end{equation}
where $\bm{I}$ is the identity tensor. Since $\tilde{\cX}$ rotates in the same way as the original coordinates under rotation and $\bm{I}$ is rotational equivariant, the final output $\rstt$ preserves the desired equivariance. The optimal parameters in the embedding and fitting networks are trained from the data. The detailed proof of the equivariance property of the formulation in Eq.~\eqref{eq:xex} is shown in~\ref{app:proof}.

The formulation in Eq.~\eqref{eq:xex} is inspired by the work of~\cite{sommers2020raman} in the context of molecular dynamics, which only has the first term $\tilde{\cX}^\top\cE\tilde{\cX}$ satisfying the equivariance requirements. However, here we need to add an extra term $\gamma \bm{I}$ to ensure that the network correctly represents the function space and symmetries that are unique for turbulence modeling. Specifically, these requirements are:
\begin{enumerate}[(1)]
    \item The Reynolds stress tensor has a full rank of three even for statistically two-dimensional (or even one- or zero-dimensional) mean flows, with all three diagonal components being nonzero in general. In contrast, the first part $\tilde{\cX}^\top\cE\tilde{\cX}$ alone degenerates for such flows: it only has a rank of two for two-dimensional mean flows.
    \item For two-dimensional mean flows, the out-of-plane stresses should be zero. This requirement is correctly satisfied by the first part $\tilde{\cX}^\top\cE\tilde{\cX}$.
\end{enumerate}
The additional term $\gamma \bm{I}$ enables the network to satisfy the first requirement of function space without breaking its equivariance property or the symmetries implied in the second requirement. Moreover, despite the apparent formal similarity of the formulation to the deviatoric decomposition of the Reynolds stress tensor, we note that our formulation is motivated mathematically rather than physically. Further details on the motivation and interpretation of the VCNN-e model are presented in~\ref{app:interp}.

With the three modules of the VCNN-e introduced above, the model provides us with a flexible representation of the Reynolds stress tensor based on the nonlocal information, and all the physical symmetries are faithfully guaranteed. The trainable parameters of the VCNN-e model include the network parameters in the embedding network and fitting network. We use the squared difference between the predicted stress and the ground truth as the loss function to optimize the parameters in the VCNN-e model. The details of data generation and training procedure will be introduced in the following sections, and the detailed architecture for the embedding and fitting networks are provided in~\ref{app:network-architecture}. Specifically, we choose $m=64$ embedding functions and their subset of size $m'= 4$ in $\cG^{\star}$ to extract feature matrix $\cD$. Similarly to the results reported in~\cite{zhang2018end,zhou2022frame}, we find the prediction accuracy is insensitive to $m'$ when $m'$ is much less than $m$. We set $m=64, m'= 4$ to preserve more features (mainly determined by $m$) in the invariant feature matrix and reduce the computational cost (mainly determined by $m\times m'$ as the input dimension) of the fitting network.

\section{Emulating a Reynolds stress transport equation with closure models}
\label{sec:results_synthetic}
In this section, we emulate a Reynolds stress transport equation with a frozen velocity field on the parameterized periodic hills, and take the associated solutions as the ground truth. Mathematically, the ground truth model defines an exact mapping from the mean velocity field to the Reynolds stress (if we do not consider any domain truncation error or discretization error). Such a setting enables us to focus on studying the learning performance of the proposed nonlocal closure model.

\subsection{Generation of training data}

\begin{figure}[ht]
\captionsetup[subfigure]{justification=centering}
\centering
\subfloat[baseline configuration]
{\includegraphics[width=0.42\textwidth]{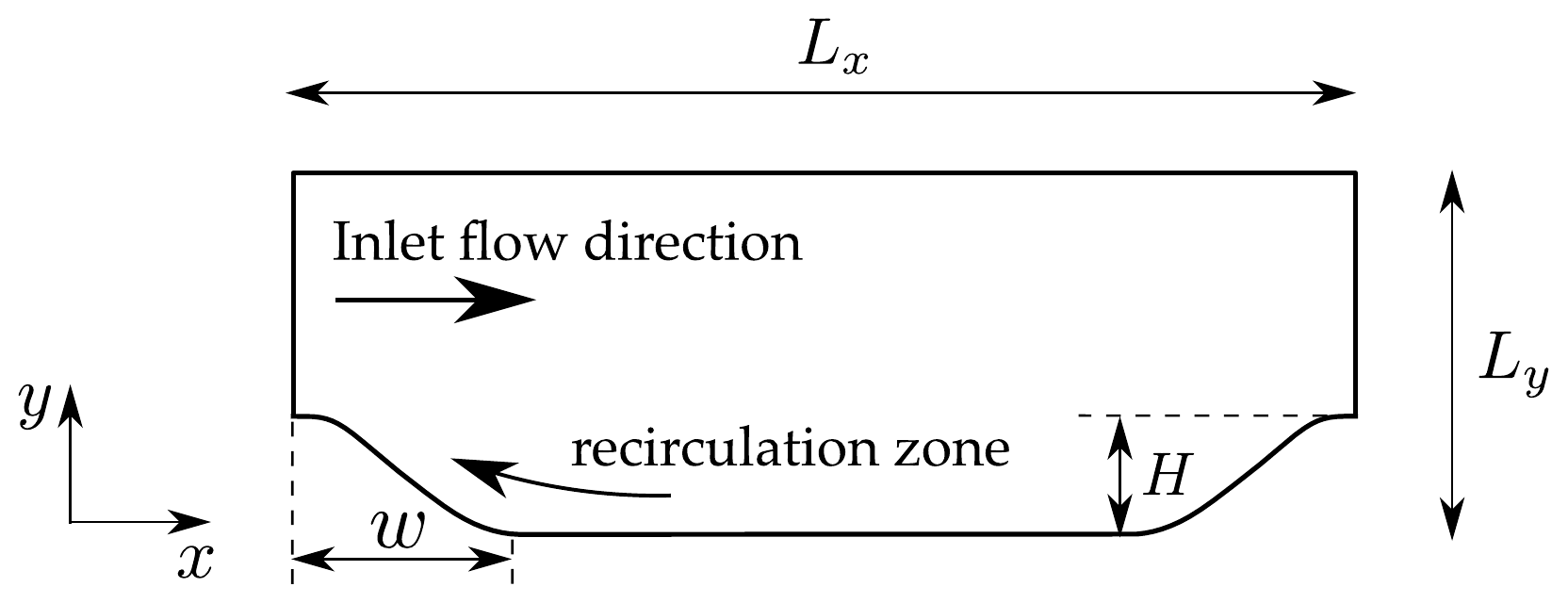}}
\hspace{0.1em}
\subfloat[configurations of varying slopes]
{\includegraphics[width=0.57\textwidth]{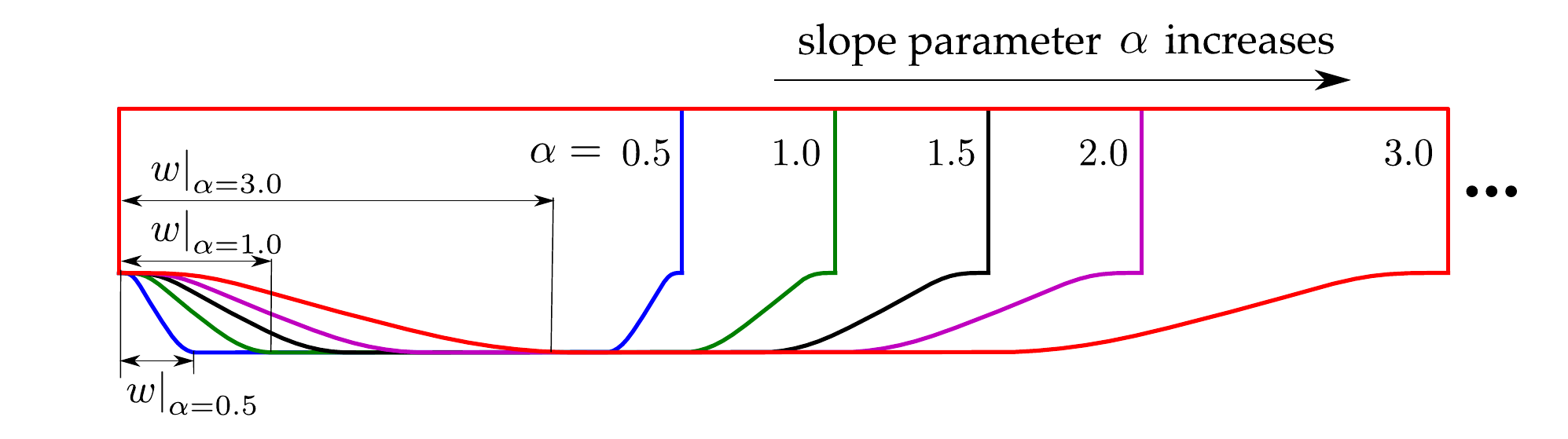}}
  \caption{Configurations of computational domains used for training and testing, showing (a) the baseline configuration in the $xy$-plane and (b) parameterized configurations with varying slopes parameters $\alpha$, defined as the ratio of the width $w$ of the hill to that of the baseline hill with $\alpha=1$ (i.e., $w|_{\alpha=1}$). The $x$- and $y$- coordinates are aligned with streamwise and wall-normal directions, respectively, and the $z$-coordinate is determined based on the right-hand rule (perpendicular to the paper and pointing outward). The slope parameters $\alpha$ are selected within the ranges $[1,\,2]$ and $[0.5,\,4]$ for training and testing flows, respectively.
  \label{fig:geo-peri-hill}
  }
\end{figure}

We consider predicting the full Reynolds stress tensor in the flows over periodic hills.
The baseline configuration of the computational domain in the $xy$-plane (normal to $z$-axis) has the length $L_x/H = 9$ and the height $L_y/H = 3.035$, both of which are normalized by the crest height $H$, which is shown in Fig.~\ref{fig:geo-peri-hill}a. The profile of the hill is described as piecewise polynomials with width $w/H = 1.93$.
We systematically vary the slope parameter $\alpha$ of the hill profile to generate a family of geometry configurations for training and testing~\cite{xiao2020flows}, as is shown in Fig.~\ref{fig:geo-peri-hill}b. The slope parameter $\alpha$ is defined as the ratio of the width $w$ of the parameterized hill to that of the baseline hill $w|_{\alpha=1.0}$. The height of the hill remains the same for all configurations, and thus the hill gets more gentle with increasing slope parameter $\alpha$. For training flows, we select 11 configurations with $\alpha = 1.0, 1.1, \ldots, 2.0$. For testing flows, the slope parameters $\alpha$ are within the range from 0.5 to 4.

We generate data by solving a Reynolds stress transport equation with the LRR-IP model (a second-moment closure model of Launder, Reece, and Rodi with isotropization of production~\cite{launder75progress}):

\begin{subequations}
\begin{equation}
    \bu \cdot \nabla \rstt
    - \nabla \cdot \left[ (\nu+\nu_t) \nabla \rstt \right]
    = \mathcal{P} + \Phi - E,
    \label{eq:rste}
\end{equation}
where the left-hand side includes convection and diffusion, and the right-hand side consists of three tensor-based terms: production $\mathcal{P}$, pressure--strain-rate $\Phi$, and dissipation tensor $E$.
The production term is closed, while the pressure--strain-rate tensor $\Phi$ consists of a slow component modeled by Rotta's return-to-isotropy model and a rapid component modeled by the isotropization of production model~\cite{pope00turbulent}. These two terms can be written as follows:
\begin{align}
    \mathcal{P} & = - \left[\mathcal{R} \cdot \nabla \bu + (\mathcal{R} \cdot \nabla \bu)^\top \right],\\
    \Phi &= \frac{C_1}{\tau} \operatorname{dev}(\rstt) + C_2 \operatorname{dev}(\mathcal{P}),
\end{align}
where $\operatorname{dev}({\cdot})$ indicates the deviatoric component of a tensor, ``$\cdot$'' denotes inner product between tensors, and $\tau = k/\varepsilon$ is the turbulence time scale. The dissipation tensor $E$ is assumed to be isotropic, i.e.,
\begin{equation}
    E = \frac{2}{3}\varepsilon \bm{I}
    \quad \text{with} \quad  \varepsilon = C_{\footnotesize{D}} \frac{k^{3/2}}{\ell_m} ,
\end{equation}
where $\bm{I}$ is the second order identity tensor, and $\varepsilon$ is the dissipation rate scalar estimated from the turbulence kinetic energy (TKE) $k$ and mixing length $\ell_m$. $k$ is defined as half the trace of the Reynolds stress: $k \equiv \frac{1}{2}\operatorname{tr}(\rstt)$, and $\ell_m$ is assumed to be proportional to wall distance $\eta$ and capped by the boundary layer thickness $\delta^*$, i.e., 
\begin{equation}
\ell_{m} = \min(\kappa \eta, C_\mu \delta^*)
\label{eq:ellm}
\end{equation}
\end{subequations}
with von Karman constant $\kappa = 0.41$. The boundary layer thickness is set to be $\delta^*/H = 0.5$ for all cases. Other model coefficients are chosen as $C_\mu = 0.09$, $C_{\footnotesize{D}} = C_\mu^{3/4}$, $C_1 = 1.8$, $C_2 = 0.6$. The boundary layer thickness is set based on the geometry for calculating the mixing length $\ell_m$ in equation~\eqref{eq:ellm}. Finally, the turbulent eddy viscosity is modeled by $\nu_t = C_\mu k^2/\varepsilon$.

We note that equations~\eqref{eq:rste}--\eqref{eq:ellm} by no means represent a complete differential Reynolds stress model since the closure terms like the diffusion and the dissipation are obtained with massively simplified models (comparable to those in incomplete one-equation models~\cite{pope00turbulent}). However, the equations capture much of the mathematical features of the Reynolds stress transport equation, including convection, diffusion, production, and dissipation. Most importantly, it includes a model for the pressure--strain-rate tensor, which is responsible for energy transfer among different components of the tensor $\rstt$. This tensor is notoriously difficult to model and is a pacing item in developing differential Reynolds stress models.
So we treat equations~\eqref{eq:rste}--\eqref{eq:ellm} as a prototype closure model for the Reynolds stress $\rstt$. Instead of modeling and solving~\eqref{eq:rste}--\eqref{eq:ellm} to get the steady solution of $\rstt$, we use VCNN-e to construct and learn a nonlocal closure model to map from the velocity field to the stress tensor directly. 

We calculate the steady Reynolds stress $\rstt(\bm{x})$ by solving the transport equation PDE \eqref{eq:rste}--\eqref{eq:ellm} with a steady, incompressible flow field $\mathbf{u}(\bm{x})$. 
The steady-state flow is driven by a constant pressure gradient such that the Reynolds number based on the volume averaged bulk velocity magnitude $u_b$, the hill height $H$, and the kinematic viscosity $\nu$ attains a specified value $Re = u_b H/\nu = 10595$. We assume the flow is statistically homogeneous in the $z$-direction so that the $z$ component of the mean velocity is 0. However, note that the instantaneous velocity fluctuations and thus the Reynolds normal stress $\mathcal{R}_{zz}$ in the $z$-direction is nonzero. The simulations of flow fields and Reynolds stress transport are performed with an open-source CFD platform OpenFOAM~\cite{opencfd21openfoam}.
Firstly, we simulate the flows over the periodic hills by solving the RANS equations with $k$--$\varepsilon$ turbulence model using the built-in steady-state incompressible flow solver simpleFoam~\cite{issa86solution}. 
Then, the simulated flow field and the corresponding Reynolds stress field are treated as the frozen flow field and initial condition, respectively, for solving the transport equation PDE \eqref{eq:rste}. When solving the RANS equations and Reynolds stress transport equation, no-slip boundary conditions are applied for the velocity at walls, wall functions are used for Reynolds stress and its associated turbulent quantities in the near-wall region, and periodic boundary conditions are applied in the streamwise direction. The numerical discretization based on the finite volume method and second-order spatial discretization scheme is utilized to solve the equations. Specifically, we make the number of cells approximately proportional to the length $L_x$ of the configuration in the $x$-direction (e.g., 200 cells for the baseline configuration with $L_x/H = 9$), while in the $y$-direction the cells are refined towards the walls and the number is fixed at 200 for all configurations. 

\begin{figure}[ht]
\centering
{\includegraphics[width=0.68\textwidth]{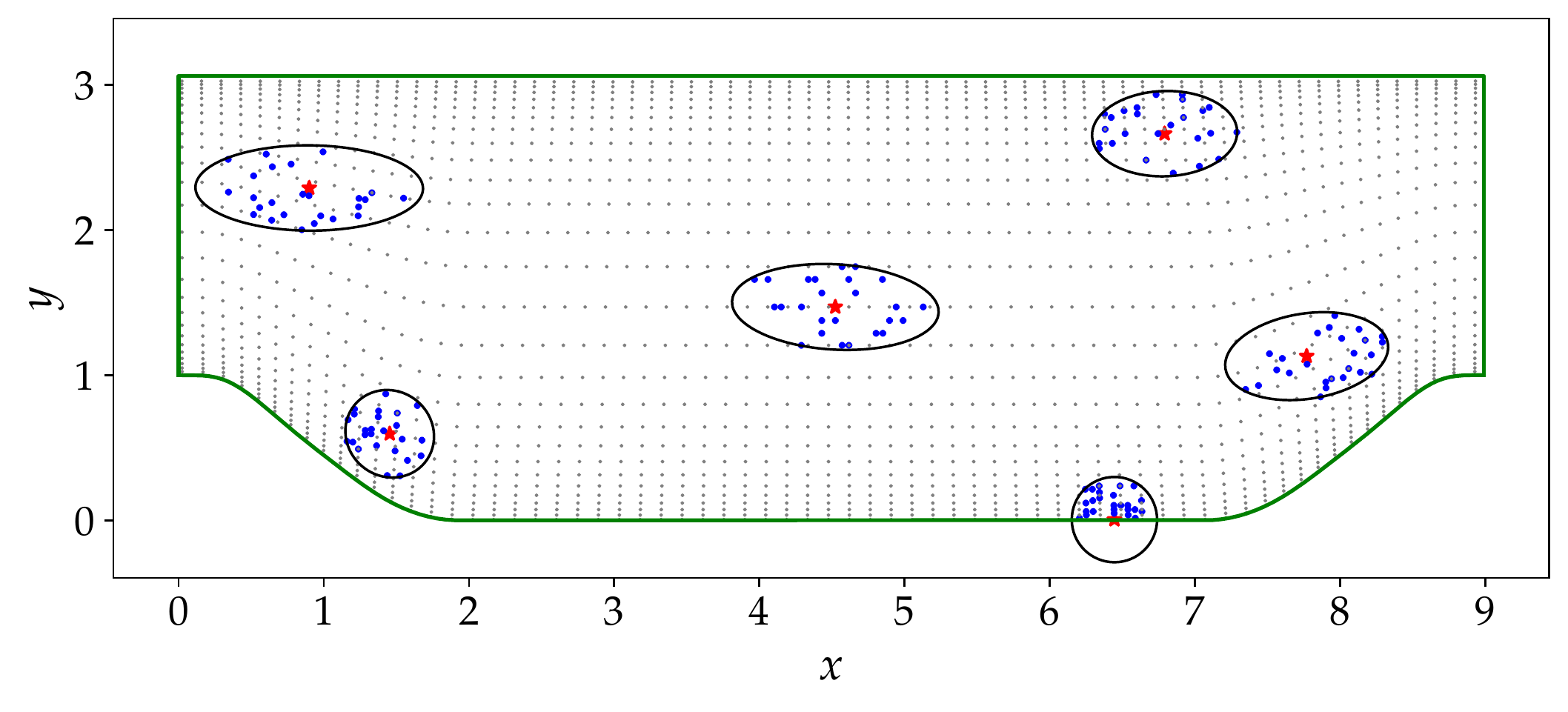}}
  \caption{Method of sampling data points within the cloud to generate pairs of labeled training data ${(\cR, \rstt)}$. The gray dots ($\color{gray!80} \bullet$) indicate all the discrete points (i.e., the cell centers) within the computational domain, showing only every seventh row and every third column for clarity. The star ($\color{red} \star$) indicates the point of interest where the Reynolds stress tensor $\rstt$ is to be predicted. The surrounding ellipse denotes the region of cloud, whose extent and orientation are determined by the velocity at the cloud center ($\color{red} \star$). The dots ($\color{blue} \bullet$) are randomly sampled within the cloud, and the set of all feature vectors attached to them is taken as the input matrix $\cQ$ to predict~$\rstt$.
  \label{fig:sampling}
  }
\end{figure}

Our aim is to learn a region-to-point mapping from a patch of nonlocal mean flow field $\region{\mathbf{u}(\bm{x})}$ to the steady Reynolds stress tensor $\rstt$ at the point of interest. As mentioned above, the nonlocal flow field is described by the input feature matrix $\cR$ based on $n$ points sampled from the cloud. The generation of such pairs of data ${(\cQ, \rstt)}$ in different locations is illustrated in Fig.~\ref{fig:sampling}.
The grey dots ($\color{gray!80} \bullet$) indicate all the discrete points (i.e., the cell centers) within the computational domain, which are shown every seventh row and every third column for clarity; the star ($\color{red} \star$) indicates the point of interest at which the Reynolds stress tensor $\rstt$ is to be predicted; $n$ data points ($\color{blue} \bullet$) are randomly sampled within the vector cloud (\tikz \draw (0,-5) ellipse (6pt and 2.5pt);) to represent the nonlocal flow field used for prediction of the Reynolds stress tensor $\rstt$. The extent and orientation of the elliptical cloud vary at different locations: The extent of the ellipse is determined by the velocity magnitude $|\bu|$ at the point of interest; the orientation (i.e., the major axis) of the ellipse aligns with the direction of the velocity $\bu$. Specifically, the length $\ell_1$ of the semi-major axis and $\ell_2$ of the semi-minor axis are defined 
according to the previous work~\cite{zhou2021learning}:
\begin{equation}
    \ell_{1} = \bigg|\frac{2 C_\nu \log\epsilon}{\sqrt{\vnorm{\bu}^2+4 C_\nu C_\zeta}-\vnorm{\bu}}\bigg| \quad \text{and} \quad \ell_{2} = \bigg|\sqrt{\frac{C_\nu}{C_\zeta}}\log\epsilon\bigg|,
    \label{eq:l}
\end{equation}
where $C_\nu$ and $C_\zeta$ are diffusion and dissipation coefficients in the 1-D steady-state convection-diffusion-reaction equation $-C_\nu \frac{\mathrm{d}^{2}}{\mathrm{~d} x^{2}} c(x) +\bm{u} \frac{\mathrm{d}}{\mathrm{d} x} c(x)+C_\zeta c(x)=P(x)$, and $\epsilon(=0.2)$ denotes relative error tolerance for truncating the associated Green's function. For the transport equation \eqref{eq:rste} in this study, the actual dissipation coefficient $C_\zeta$ in Eq.~\eqref{eq:l} should be approximately $\frac{2}{3} C_{D} \frac{k^{1 / 2}}{\ell_{m}}$, considering the dissipation term $\mathcal{E}$. Here, it is assumed as a constant $C_\zeta \approx 2$ to determine the extent of the clouds, where the maximum $\ell_m$ and $k$ are used for the approximation. Similarly, the actual diffusion coefficient $C_\nu$ in Eq.~\eqref{eq:l} is chosen as ${(\nu_{t}+\nu)}_{\min} \approx 0.01$.

After determining the extent of clouds, we randomly sample $n=300$ data points within each cloud to generate training data. The available data points are repeatedly sampled for the locations where the number of data points in the ellipse is smaller than 300. It should be noted that the proposed neural network is flexible with an arbitrary number of sampled points in the cloud. Here we set $n$ as a constant for training in order to conveniently store the data and process them in batches for better computational efficiency during training. 
There are $6.6 \times 10^5$ pairs of ${(\cR, \rstt)}$ in total for 11 training flows with $\alpha = 1.0, 1.1, \ldots, 2.0$. Considering that the adjacent data pairs from the same flow may appear similar, we take one from every two adjacent pairs to reduce the size of the training dataset and use the reduced $3.3 \times 10^5$ pairs of ${(\cR, \rstt)}$ as the training data. As mentioned above, the data are non-dimensionalized by the characteristic length $H$, flow velocity $u_b$, and time $H/u_b$, and then fed into the neural network.

\subsection{Neural-network-based prediction of Reynolds stress tensors}
\label{sec:res-tau}
With the generated training data described above, one can directly train the VCNN-e model in a straightforward way. However, we employ a different progressive training strategy here to boost training efficiency. The network is initially trained using sparse data (i.e., a small stencil size), followed by dense data. This training method achieves better performance due to the accelerated optimization process compared to the conventional training method, which is illustrated in~\ref{app:training-method-compare}.
The training processes are performed on an NVIDIA RTX 3090 GPU using the open-source machine learning framework PyTorch~\cite{paszke2019pytorch}. The code for data generation and model training are available on GitHub~\cite{zhou2022vcnne-git}, which can be used by the readers for reproducing the results and further development.

After training, the prediction capabilities of the trained neural-network-based model are investigated in different configurations with slope parameters $\alpha$ between 0.5 and 4. In the testing step, we take all the data points within the clouds (i.e., full stencil size varying from 50 to 2000 points) to predict the Reynolds stress tensor at the cloud centers, which is different from using only 300 sampled points in each cloud for prediction in the training step. In this sense, testing on the same flow used for training is still compelling because the input are different.
We examine how well the learned model performs toward capturing two turbulent quantities: (1) turbulence kinetic energy (TKE) $k$ and (2) Reynolds stress anisotropy invariant $A_2$ (detailed below), both of which are invariants of Reynolds stress tensor. Note that the loss function for training the network only consists of the differences in Reynolds stress components,
while the invariant quantities, such as the TKE and the anisotropy scalar, are not included. Accurate predictions of these invariant quantities rely on the equivariant property of the proposed VCNN-e model.

\begin{figure}[ht]
\centering
{\includegraphics[width=0.9\linewidth]{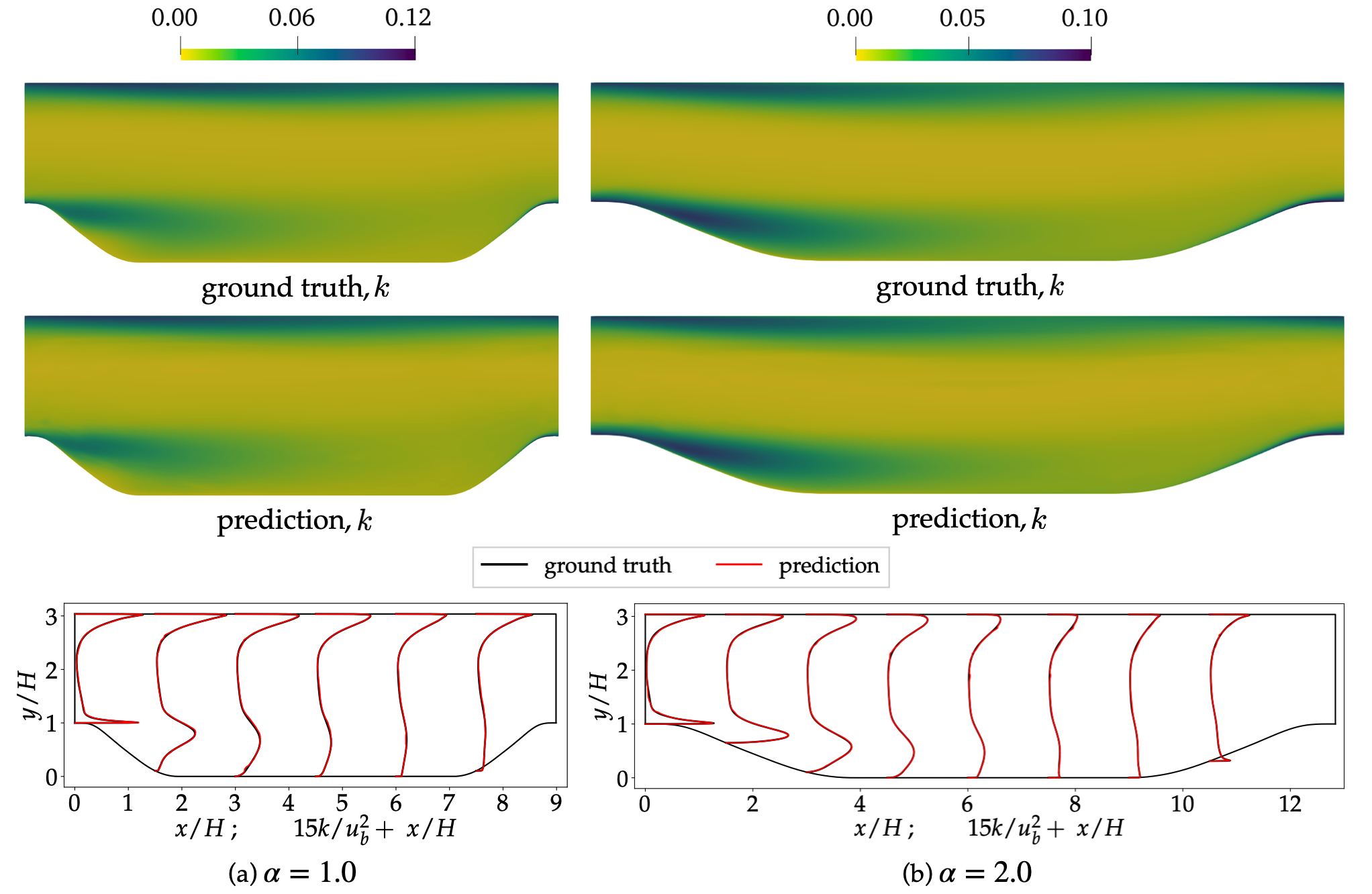}}
  \caption{Comparison of the ground truths of the \textbf{turbulence kinetic energy} (TKE) $k$ (top row) and the corresponding predictions based on the predicted Reynolds stress from the trained neural network (middle row), along with the TKE profiles at six and eight cross-sections (bottom row) for two  \textbf{interpolated} configurations with slope parameters $\alpha=1$ (left panels) and $\alpha =2$ (right panels).
  }
  \label{fig:TKE-interpolation-compare}
\end{figure}

The predicted TKE $k$ fields are nearly identical to the corresponding ground truths in two extreme interpolated flows with $\alpha=1$ and 2, which is illustrated in Fig.~\ref{fig:TKE-interpolation-compare}. The TKE is calculated based on the predicted Reynolds stress and then compared with the ground truth. We can observe similar patterns between the predicted TKE and the ground truths (top two rows). The similarity is shown more clearly at the vertical cross-sections (bottom row): The predicted TKE profiles coincide with those of the corresponding ground truths. Such good prediction performance in the interpolated flows lies in the fact that the method respects the transport physics of turbulent quantities by considering the mean flow field in a nonlocal region. Moreover, 300 randomly sampled points in each cloud are sufficient to describe the flow, allowing the network to learn the nonlocal mapping and make accurate predictions.

\begin{figure*}[!htb]
\centering
{\includegraphics[width=0.9\linewidth]{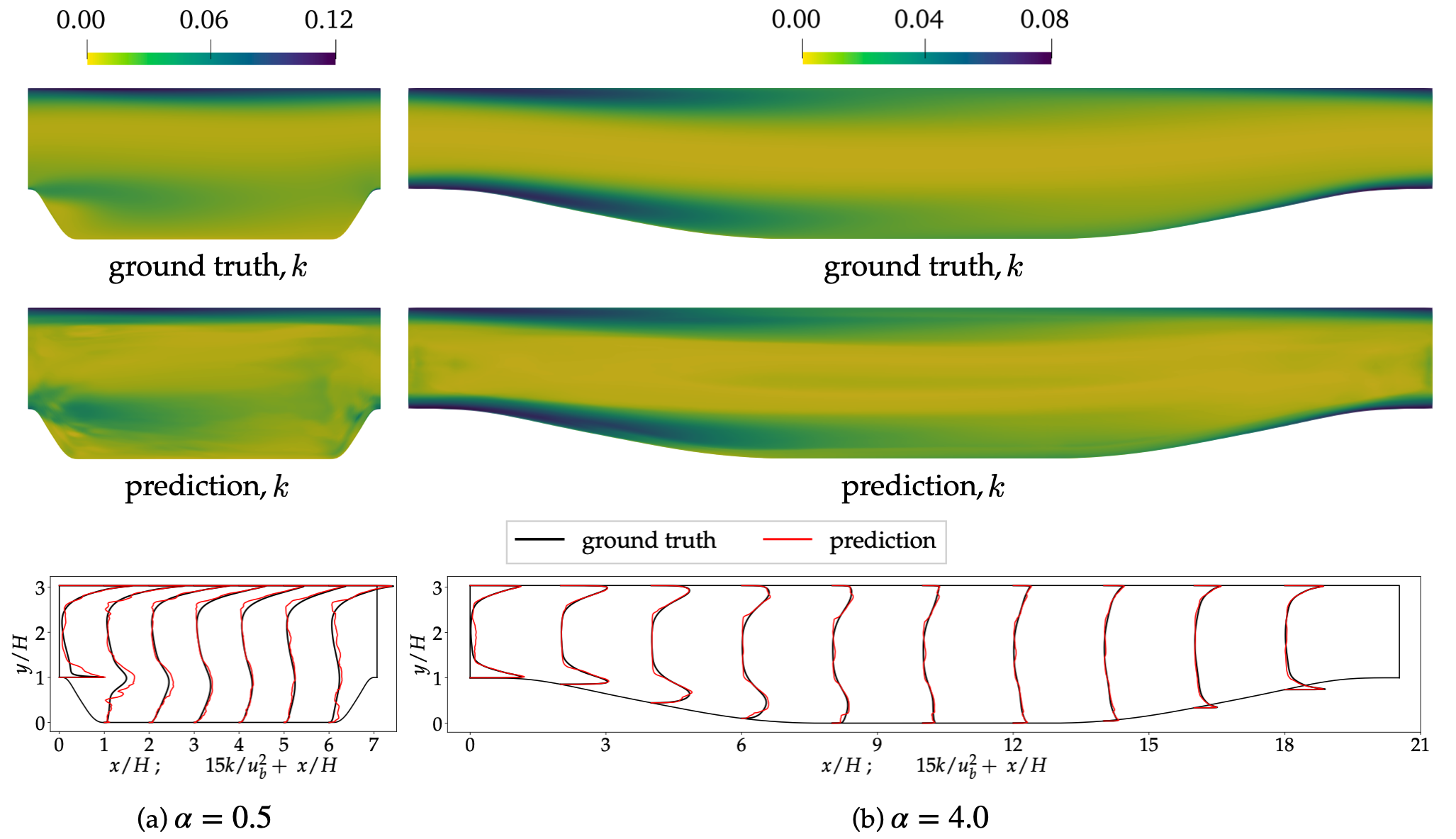}}
  \caption{Comparison of the ground truths of the \textbf{turbulence kinetic energy} $k$ (top row) and the corresponding predictions based on the predicted Reynolds stress from the trained neural network (middle row), along with the TKE profiles at seven and ten cross-sections (bottom row) for two extreme \textbf{extrapolated} configurations with slope parameters $\alpha=0.5$ (left panels) and $\alpha =4.0$ (right panels).
  }
  \label{fig:TKE-extrapolation-compare}
\end{figure*}

The prediction of TKE for two extreme extrapolated flows with $\alpha=0.5$ and 4 is not as accurate as that for the interpolated flows, but is still reasonable with expected features being captured. As illustrated in Fig.~\ref{fig:TKE-extrapolation-compare}, the nonlocal model is able to predict the TKE in the boundary layer at the top wall for both flows. However, despite some similarity with the ground truth in the near-hill region, the flow with $\alpha=0.5$ exhibits distinct inconsistency and artifacts due to the steepest hill slope and the large flow separation. In contrast, the flow with $\alpha=4.0$ shows a better prediction near the bottom wall because of the gentlest hill slope. The comparison is more clearly illustrated in the profile plots at cross-sections (bottom row). The predicted TKE in the flow with $\alpha = 0.5$ is overestimated on the leeside of the hill, whereas the predicted TKE in the flow with $\alpha = 4.0$ is nearly identical to the ground truth despite some unsmoothness.

Another evaluation criterion of predictive capability is the Reynolds stress anisotropy invariant $A_2$ as in practical flows the Reynolds stress tensor is anisotropic due to deformation of large eddies by mean strain, inhomogeneities and boundaries. The invariant $A_2$ is defined as half the square of the tensor magnitude of normalized Reynolds stress anisotropy $\bm{b}$ given by~\cite{lumley1977return}:
\begin{equation}
A_2 = b_{i j} b_{j i}/2 , \quad \text { with} \quad b_{i j}=\frac{\rstt_{i j}}{2 k}-\frac{1}{3} \delta_{i j},
\end{equation}
where $k$ is the TKE and $\delta_{i j}$ is the Kronecker delta. Here we scale up the invariant $A_2$ by multiplying the TKE $k$, considering that $A_2$ of anisotropy tensor is no longer important when $k$ is sufficiently small. The predicted $k A_2$ are then compared with the corresponding ground truths in both interpolated and extrapolated flows. 

\begin{figure*}[t]
\centering
{\includegraphics[width=0.9\linewidth]{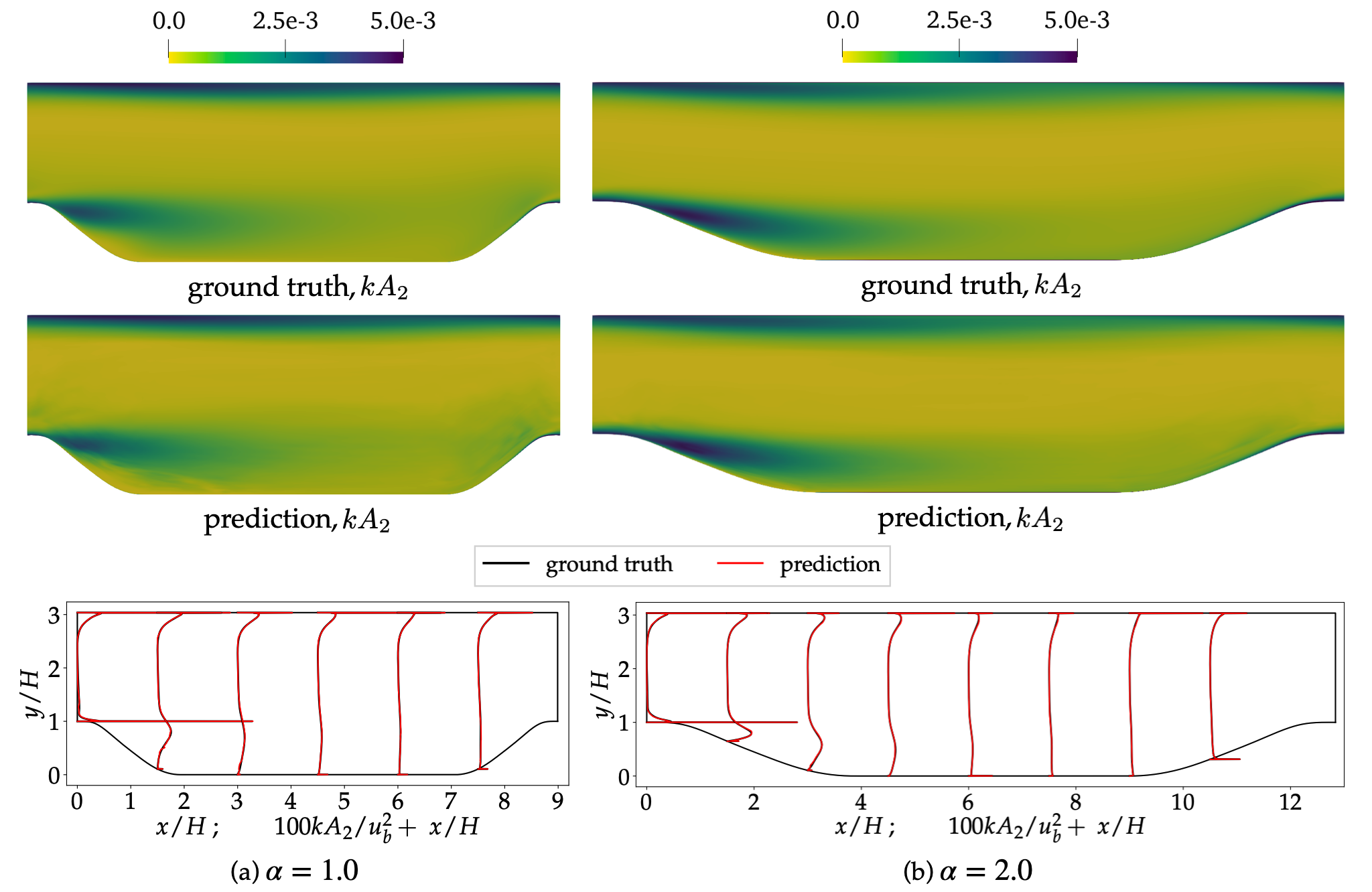}}
  \caption{Comparison of the ground truths of $kA_2$, the \textbf{Reynolds stress anisotropy invariant} up scaled by the turbulent kinetic energy (top row) and the corresponding predictions based on the predicted Reynolds stress from the trained neural network (middle row), along with the $k A_2$ profiles at several cross-sections (bottom row) for two \textbf{interpolated} configurations with slope parameters $\alpha=1$ (left panels) and $\alpha =2$ (right panels).
  }
  \label{fig:A2-interpolation-compare}
\end{figure*}

Similarly, the predicted $k A_2$ in two extreme interpolated flows with $\alpha=1$ and 2 are quite accurate with nearly no deviation from the ground truths, as is shown in Fig.~\ref{fig:A2-interpolation-compare}. The nonlocal model is able to capture a more rapid change of $A_2$ near walls compared to that of TKE. For extrapolated flows, however, the situation is different since the anisotropy invariant $A_2$ depends sensitively on strain and rotation rate changes and may vary dramatically as the hill's shape changes. For clarity, we have shown the prediction results for flows with $\alpha=0.5$ and 4 in Fig.~\ref{fig:A2-extrapolation-compare} in~\ref{app:extra-kA2}. 

\begin{figure}[htb]
\centering
\includegraphics[width=0.9\linewidth]{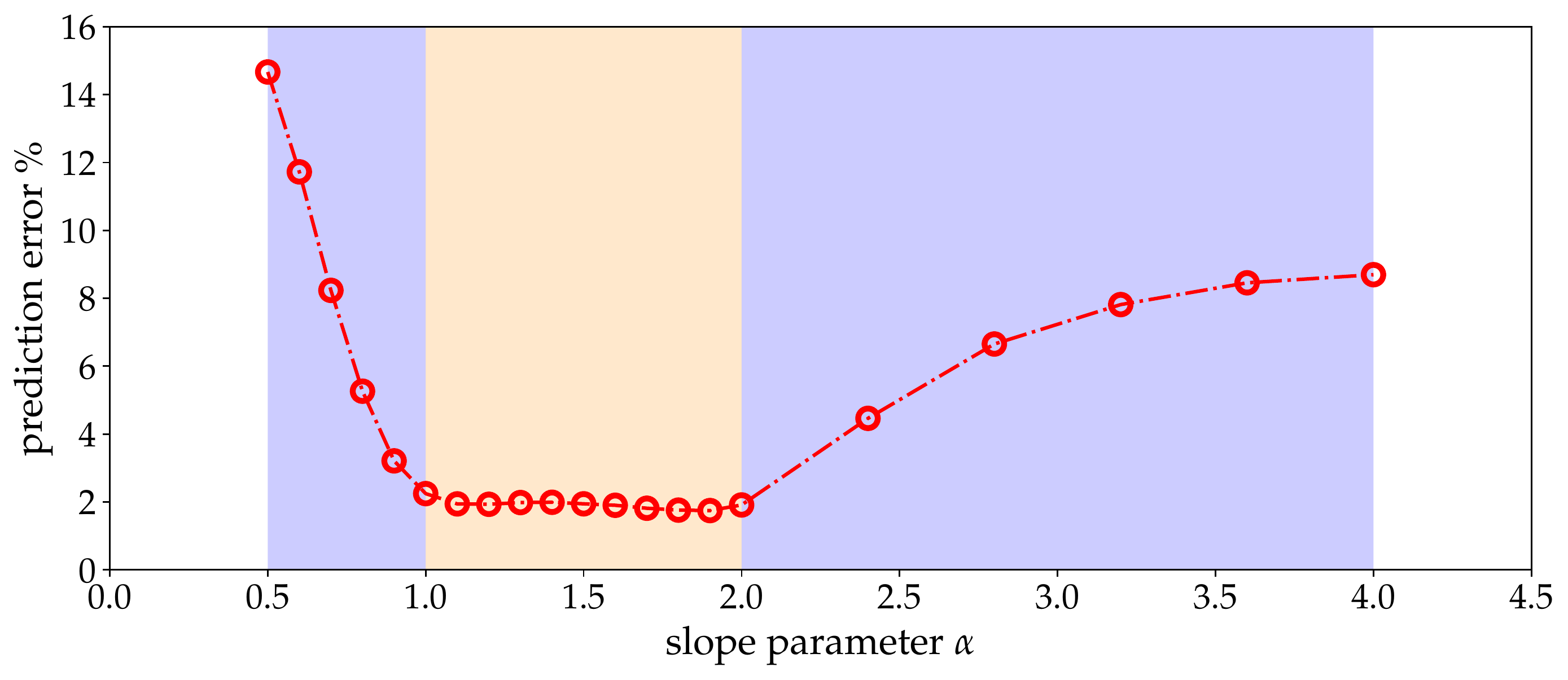}
  \caption{Predictive errors for turbulence kinetic energy $k$ at various slope parameters $\alpha$. The yellow/lighter and blue/darker backgrounds represent the regimes of the slope parameters $\alpha$ of the interpolated and extrapolated testing flows, respectively. The neural network is trained with data of a fixed stencil size from 11 flows with $\alpha = 1, 1.1, \ldots, 2$.
  The trained network is then tested on the data of full stencil size from the same 11 interpolated flows and 10 extrapolated flows with $\alpha = 0.5, 0.6, \ldots, 0.9$ and $\alpha = 2.4, 2.8, \ldots, 4$.
  \label{fig:prediction-error-TKE}
  }
\end{figure}

From the above results, we can see that the hill slope significantly impacts the prediction performance of the trained model. For further investigation, we evaluate the generalizability of the trained model on 21 configurations with varying slope parameters $\alpha$ from 0.5 to 4. The prediction performance is assessed using the prediction error, which is defined as the normalized $\ell^{2}$-norm discrepancy between the calculated TKE $\hat{k}$ based on the predicted Reynolds stress and corresponding ground truth $k^*$:
\begin{equation}
    \text{error} = \frac{\sqrt{\sum_{i=1}^{N} {|\hat{k}_i -
    k_{i}^*|}^2}}{\sqrt{\sum_{i=1}^{N} {|k_{i}^*|}^2}}, 
    \label{eq:error-def}
\end{equation}
where the summation is performed on all of the $N$ training or testing data points (e.g., 40000 data points for baseline configuration with $\alpha=1$). Again, the prediction is made using the full stencil size, which differs from using a  fixed number of sampled points in the training step, so the flows with $\alpha$ between 1 and 2 can still serve as the interpolated testing flows. The prediction errors for all 21 testing flows are illustrated in Fig.~\ref{fig:prediction-error-TKE}. The middle (yellow/light gray) region from 1 to 2 represents the regime of slope parameters $\alpha$ of training flows while that on both sides (blue/dark) represents the regime of extrapolated testing flows. The prediction performance on 11 interpolated testing flows is the best, with prediction errors being 1.8\%, showing the trained model can accurately predict the Reynolds stress tensor based on a nonlocal flow field. When extrapolating to flows with steeper ($\alpha$ \textless\;1) or more gentle ($\alpha$ \textgreater\;2) hill profiles, the prediction errors increase significantly to 14.7\% for $\alpha=0.5$ and 8.7\% for $\alpha=4$, which is reasonable and can be partially explained by the changes of input and output distributions displayed subsequently.

\begin{figure}[!htb]
\centering
{\includegraphics[width=0.99\textwidth]{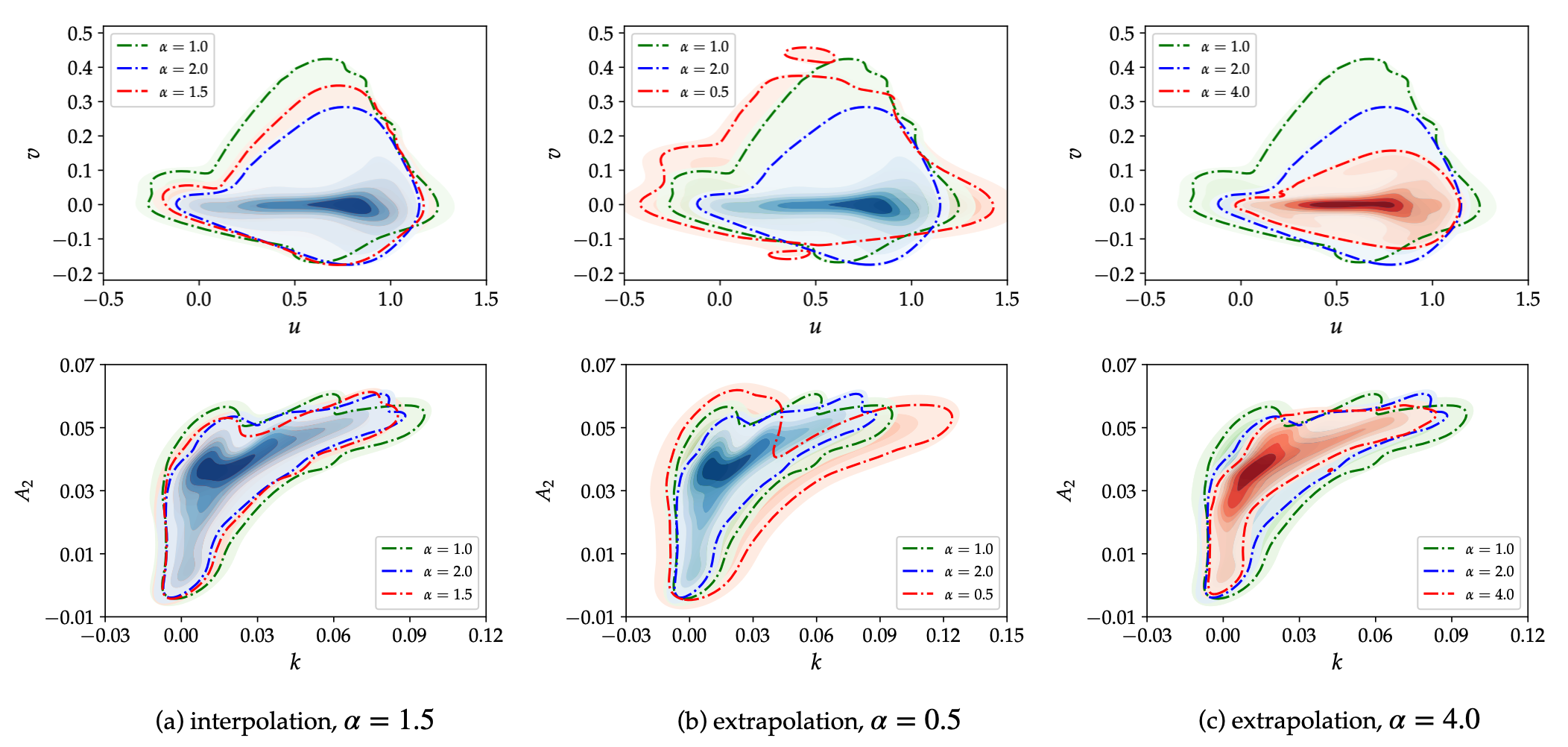}}
  \caption{First row: kernel density estimate (KDE) plot of the joint data distribution of the velocities in $x/y$ directions $(u/v)$ with different slope parameters $\alpha$, in the regimes of interpolation and extrapolation on two sides. Second row: KDE of the joint data distribution of the turbulence kinetic energy $k$ and anisotropy invariant $A_2$ with different slope parameters $\alpha$, in the regimes of interpolation and extrapolation on two sides. The dash-dotted line represents the contour line which envelops 99\% of the probability mass of the data.
  \label{fig:data-distribution}
  }
\end{figure}

We visualize the input and output data distributions in two directions of extrapolation through the kernel density plot of the variable pairs with different $\alpha$ in Fig.~\ref{fig:data-distribution}: velocities ($u/v$) in $x/y$-directions in the first row and turbulence kinetic energy $k$/anisotrioy invariant $A_2$ in the second row. We can see that in subfigure (a) (the first column), the interpolation regime, both the input data $u,v$ and output-dependent data $k, A_2$ have similar distributions as $\alpha$ change values from 1 to 1.5, and to 2. In subfigure (c) (the last column), the data distribution corresponding to $\alpha=4$ keeps the trend of change and deviates from the interpolation data a little bit. In contrast, in subfigure (b) (the middle column), the data distribution corresponding to $\alpha=0.5$ presents some data points that are more significantly away from the training data.

\section{Results on a dataset from direct numerical simulations}
\label{sec:results_dns}

In this section, we consider using the mean flow fields collected from direct numerical simulations to predict the corresponding Reynolds stress. Compared to the setting in Section~\ref{sec:results_synthetic}, this setting is much closer to the practice of developing and calibrating closure models and hence provides a more realistic testbed for the proposed nonlocal closure model. However, it should be noted, due to the information loss in the Reynolds averaging, there may not exist an exact mapping from the mean flow fields to the Reynolds stress. This is evident from the exact transport equations for the Reynolds stresses~\cite{launder1975progress}, where the turbulent diffusion depends on the triple correlation of the turbulent fluctuations while the pressure-strain-rate term also depends on the pressure fluctuations. However, in RANS models such information is not available and the corresponding term must be modeled as functions of the mean field. Using only the mean flow fields as the input without turbulent fluctuation data can result in modeling inaccuracy. So the objective here is to investigate the prediction performance of the proposed nonlocal model in the presence of such a modeling error.

\subsection{Dataset for nonlocal turbulence modeling from direct numerical simulations}

A new dataset of flows over parameterized periodic hills from direct numerical simulations (DNS) is employed for the further evaluation of VCNN-e in modeling the Reynolds stress. Direct numerical simulations are performed on 29 parameterized periodic hills with five different hill slopes ($\alpha = 0.5, 0.75, 1.0, 1.25, 1.5$), three different domain lengths ($L_x/H = 3.858 \alpha+2.142, 3.858 \alpha+5.142, 3.858 \alpha+8.142$) and heights ($L_y/H = 2.024, 3.036, 4.048$). In this study, we select five flows with varying hill slopes for training and testing, in which their domain lengths and height are $L_x/H = 3.858 \alpha+5.142$ and $L_y/H = 3.036$, respectively. The DNS data is generated by Sylvain Laizet and his coworker using the high-order flow solver {\fontfamily{lmtt}\selectfont Incompact3d}~\cite{xiao2020flows,xiaoperiodichill-git,laizet2011incompact3d}, which consists of the mean velocity, pressure, and Reynolds stress fields. The DNS data is then mapped to their counterparts on a coarser mesh to generate the training and testing data in this situation. For example, in flow with $\alpha=0.5$, the DNS data on fine mesh ($640 \times 385 \times 192$) is first averaged in the spanwise direction ($z$) and then interpolated to the data on a coarse mesh ($157 \times 150$).

After the data preprocessing, the generation of the training and testing data follows a similar method to that in Section~\ref{sec:results_synthetic}. Specifically, the cloud extent is based on the local velocity and determined by Eq.~\eqref{eq:l} with the numerical values of $\epsilon$, $C_\nu$, and $C_\zeta$ being 0.2, 0.02, and 2, respectively; the cloud orientation (i.e., long axis) aligns with the local velocity. 
A constant stencil size of 300 is applied for sampling in each cloud. In this section, four different flows with $\alpha = 0.5, 0.75, 1.25, 1.5$ are used for training and the flow with $\alpha = 1.0$ for testing. Finally, we have $1.2 \times 10^5$ pairs of ${(\cR, \rstt)}$ as the training data and $3 \times 10^4$ pairs as the testing data. Considering the significant reduction in the amount of training data compared to the synthetic situation, the VCNN-e is trained straightforwardly rather than in a progressive way.

\subsection{Neural-network-based prediction of Reynolds stress tensors}

\begin{figure}[t]
\centering
\includegraphics[width=0.99\linewidth]{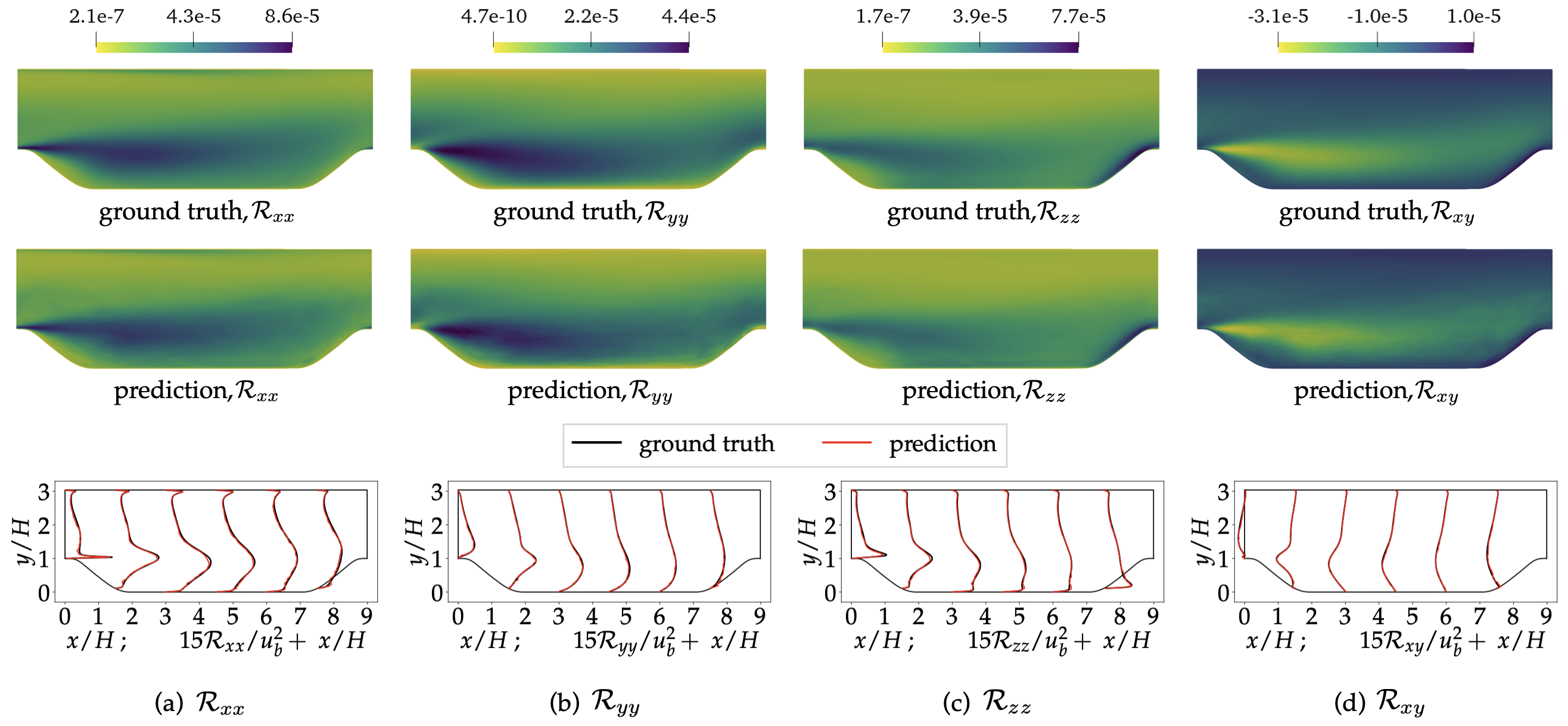}
\caption{Comparison of the ground truths (DNS) of four Reynolds stress components (top row) and the corresponding predictions provided by the trained neural network (middle row), along with their scaled profiles at six cross-sections (bottom row) for the testing configuration with the slope parameter $\alpha = 1.0$.}
\label{fig:DNS-compare}
\end{figure}

The predicted Reynolds stress in the testing flow exhibits a high degree of agreement with the ground truth. Comparison of the predicted Reynolds stress components ($\cO_{xx}$, $\cO_{yy}$, $\cO_{zz}$, and $\cO_{xy}$) and their corresponding ground truths is shown in Fig.~\ref{fig:DNS-compare}. The predictions (middle row) of four components exhibit similar patterns to the ground truths (top row). The similarity is further illustrated by the consistency of their profiles on six cross-sections ($x/H = 0, 1.5, \cdots, 7.5$). Such consistency further demonstrates the capability of VCNN-e to model the Reynolds stress in a more realistic and complex scenario in which the transport is driven by the original Navier-Stokes equations rather than an explicit transport PDE. Note that, when working with a synthetic PDE in the former section, there is an exact mapping between mean flow fields and Reynolds stress, but this may not be the case for the DNS situation. The consistency is further demonstrated by the predicted turbulent anisotropy states in Fig.~\ref{fig:DNS-triangle} in~\ref{app:anisotropy}. The prediction over three cross-sections exhibits a much more similar distribution to the DNS data in the Barycentric triangle than the RANS result.

\begin{figure}[t]
\centering
\includegraphics[width=0.99\linewidth]{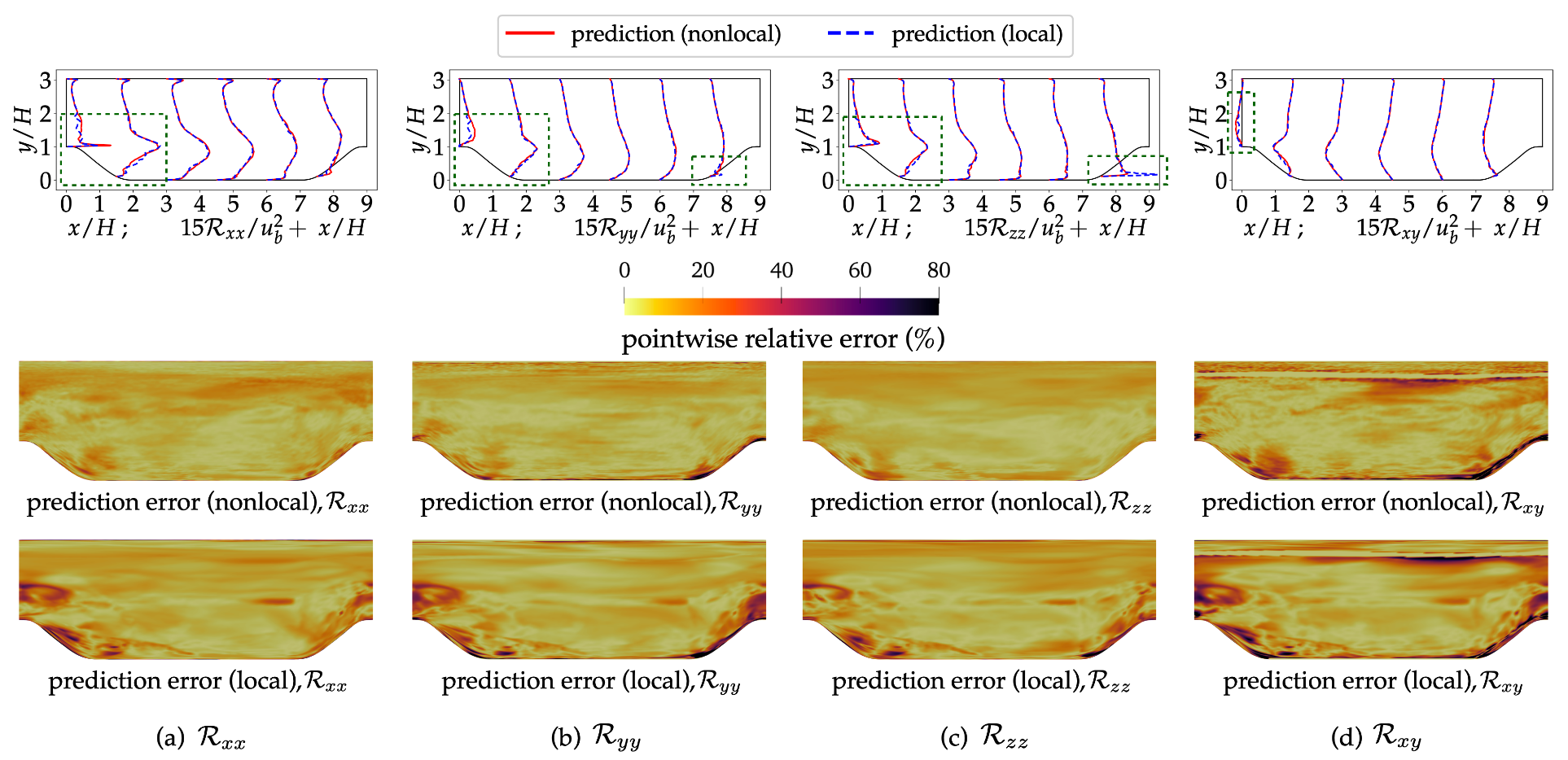}
\caption{Comparison of the predicted  Reynolds stress components ($\cO_{xx}$, $\cO_{yy}$, $\cO_{zz}$, and $\cO_{xy}$) from the trained nonlocal and local models: Scaled profiles of the predicted Reynolds stress components from the nonlocal (solid/red) and local (dashed/blue) models are compared at six cross-sections (top row), with dashed boxes indicating locations with distinct differences; pointwise prediction errors of Reynolds stress components from the nonlocal (middle row) and local (bottom row) models, with the dark area representing a larger deviation from the ground truths. 
\label{fig:DNS-local-nonlocal}}
\end{figure}

Given the success of VCNN-e in predicting the Reynolds stress in the DNS dataset, we further examine the role of nonlocality in the modeling. To this end, we train a network using only local input features and then compare the local model with the nonlocal counterpart. In particular, the local model is trained with five scalar features $\boldsymbol{c}=\left[\mathrm{u}, s, b, d, p\right]^{\top}$, whereas features associated with neighboring points are removed. The training settings are kept the same to ensure a fair comparison. The prediction error from the local model (14.9\%) is significantly larger than that from the nonlocal model (7.7\%), indicating a greater deviation from the ground truth compared to the nonlocal model. Note that the calculation of the prediction error here follows Eq.~\eqref{eq:error-def} but targets at the total squared error of all the components of the Reynolds stress tensor instead of the TKE. The comparison of the predictions from the local and nonlocal models is shown in Fig.~\ref{fig:DNS-local-nonlocal}. Specifically, the profiles of four Reynolds stress components from the two models are compared at six cross-sections (top row), with the green/dashed box designating regions with clear differences. We can observe that the regions with differences are primarily found close to the hills, whereas in other areas the predictions from the two models are almost similar. The error plots (bottom two rows) illustrate this phenomenon more clearly, which show the pointwise relative error compared to the ground truth, with the darker area indicating a larger prediction error. For all Reynolds stress components, it is evident that the nonlocal model provides more accurate predictions than the local model near the inlet/outlet and in the recirculation zone. This is because the contraction and expansion of the flow channel induce rapid changes in strain and rotation, the history of which influences the Reynolds stress anisotropy downstream. That is to say, the Reynolds stress anisotropy in the windward and leeward sides near the hill cannot be determined by using only local flow features; the upstream flow field must also be considered. Our nonlocal model can capture such nonlocality and hence yields a more accurate prediction of the Reynolds stress than the local model.

\begin{figure}[t]
\centering
\includegraphics[width=0.99\linewidth]{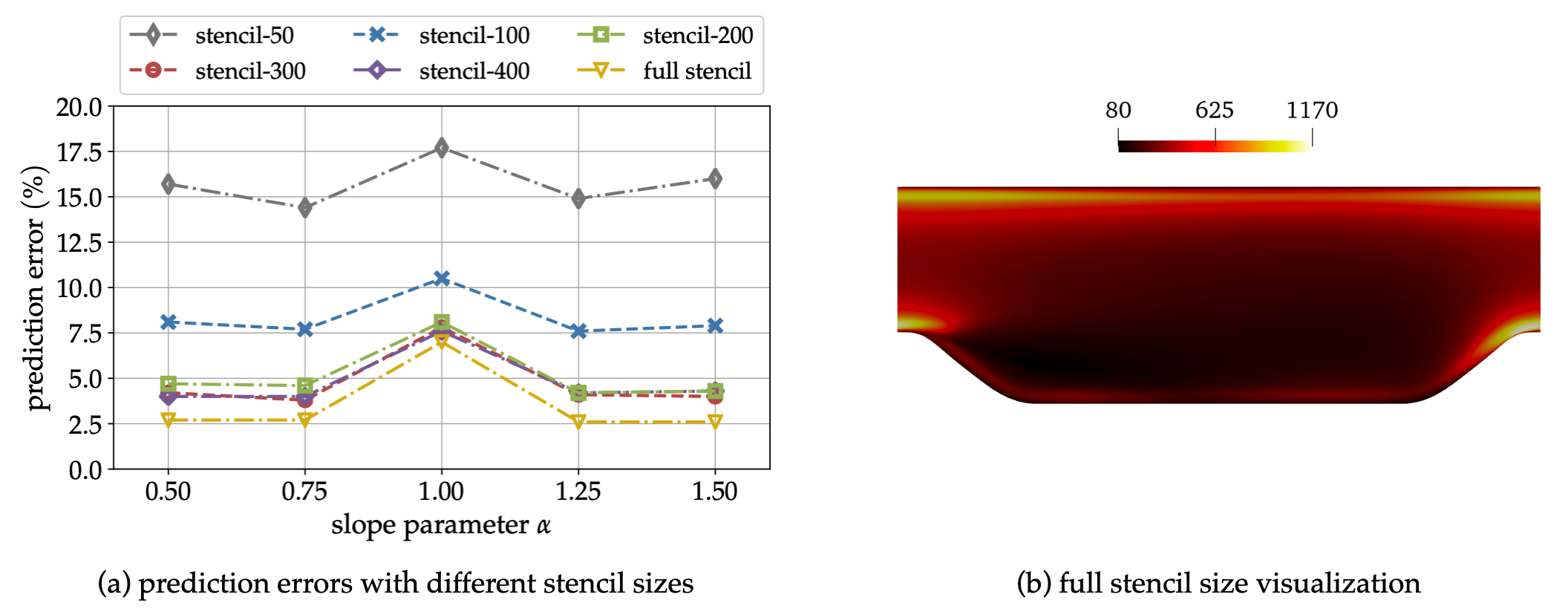}
\caption{Resolution adaptivity property of VCNN-e: (a) prediction errors for five periodic-hill flows with different stencil sizes and (b) Visualization of full stencil size at different locations in the testing flow with slope parameter $\alpha=1.0$. Full stencil indicates that all the data points in the cloud are used for predicting the Reynolds stress at the central location, which differs from other testings with fixed stencil sizes (varying from 50 to 400) in that the stencil size varies from location to location.}
\label{fig:reso-inva}
\end{figure}

Furthermore, the capability of VCNN-e to handle different stencil sizes demonstrates its resolution adaptivity and that it can serve as a neural operator for developing nonlocal tensorial constitutive models, which is illustrated in Fig.~\ref{fig:reso-inva}.
To be specific, the network is trained using a constant stencil size of 300 and tested with different stencil sizes ranging from 50, 100, 200, 300, and 400 up to a full stencil size that varies by location. Notably, we specify separate seed states in the code to perform different random sampling for different stencil sizes to ensure that the testing data are unique from the training data and that the testing data for larger stencil sizes do not contain those for smaller stencil sizes. The result indicates that the prediction errors gradually converge as the stencil size increases. Comparatively, the prediction from the stencil size of 50 is inferior to that from other stencil sizes because 50 randomly sampled points are insufficient to describe the nonlocal flow field in the cloud, resulting in a considerable loss of mean flow information. In contrast, the full stencil size achieves the best prediction performance since all of the data points inside the cloud are used to represent the nonlocal mean flow fields without sacrificing any flow information. The distribution of the full stencil size over the entire domain is visualized in Fig.~\ref{fig:reso-inva}(b) to highlight its variation at different locations, with the dark area indicating a small stencil size. The prediction errors for the intermediate stencil sizes, from 200 to 400, are fairly close to those for the full stencil size. The above result demonstrates that the data with varying stencil sizes (i.e., neural network input dimension) can share the same network parameters, and all of them provide Reynolds stress predictions close to the DNS data. In other words, the VCNN-e is capable of learning the operator that maps an arbitrarily discretized nonlocal mean flow field to the Reynolds stress.

\section{Discussions}

So far we have mainly demonstrated the capability of VCNN-e in modeling the full Reynolds stress. The same methodology can also be straightforwardly applied to model the normalized Reynolds stress anisotropy, which can be derived from the Reynolds stress transport equation~\eqref{eq:rstm} and implies a similar region-to-point mapping as the latter does. It is well known that turbulence modeling consists of determining the length scale, velocity scale, and anisotropy of the turbulence. The former two quantities amount to eddy viscosity $\nu_t$, which cannot be determined by the local mean flow (i.e., the mean velocity or its gradient). 
Therefore, turbulent length- and velocity-scales must be solved for, i.e., via transport PDEs as in $k$--$\varepsilon$~\cite{launder1974application} and Spalart--Allmaras models~\cite{spalart92one}. These models account for the nonlocal dependence of eddy viscosity on the mean flow fields. However, they do not account for the dependency of the Reynolds stress \emph{anisotropy} on the mean flow upstream. Rather, the normalized anisotropy is modeled based on local mean velocity gradient (or equivalently $\nabla \mathbf{u} = \bm{S} + \bm{\Omega}$, with $\bm{S}$ and $\bm{\Omega}$ denoting strain-rate and rotation-rate tensors, respectively). As such, eddy viscosity models (linear or nonlinear) are referred to as \emph{weak equilibrium} models~\cite{pope1975more}.
Modeling the normalized Reynolds stress anisotropy through the proposed VCNN-e can complement the existing eddy viscosity models by providing a better description of the Reynolds stresses.

Additionally, it is important to compare new methods with existing methods and to highlight actual improvement and innovation. We note that one machine learning model potentially competitive for predicting Reynolds stress or other tasks in CFD is the graphical kernel network (GKN)~\cite{li2020neural}. It has permutational invariance built-in and can handle unstructured data, which is assumed in the present work and very common in CFD. However, GKN in itself does not have built-in rotational invariance, but one can equip GKN with rotational invariance by pre-processing the input. Therefore, it is valuable to compare VCNN to GKN, without and with rotational invariance. Our recent work~\cite{zafar2021frame} has performed a comprehensive comparison. The findings can be summarized as follows. 
(1) GKN without rotational invariance performs poorly when predictions are made in rotated coordinates.
(2) GKN equipped with rotational invariance has slightly better accuracy than VCNN.
(3) The training time of GKN scales quadratically with the number of points in the cloud, i.e., $\mathcal{O}(n^2)$, where $n$ is the number of points in the cloud. Such computation scaling makes it difficult for GKN to handle clouds with over 200 points. In contrast, VCNN has a constant training time (due to parallel processing of the GPU), i.e., $\mathcal{O}(1)$. It can handle clouds with at least a few thousand points.
The comparison between GKN and VCNN discussed above is performed on a scalar transport equation, but the conclusion for a tensor transport equation should be qualitatively similar.

\section{Conclusion}
In this work we generalize the vector-cloud neural network as a neural operator to model the constitutive tensor transport equations. By training on unstructured data points, the proposed neural operator can faithfully capture the underlying nonlocal physics through a region-to-point mapping; it is invariant to coordinate translation and ordering of the points and meanwhile equivariant to coordinate rotation.
The demonstrated performance shows its promise for nonlocal constitutive models, especially the RANS momentum equation in turbulence modeling to solve for mean velocities and pressure. 

There are a few directions worth exploring in future work.
The immediate next step is to examine the performance of the learned nonlocal model as the closure model for the primary PDEs. The recent work~\cite{xu2022pde} has demonstrated the robustness and stability of the neural operator-based eddy viscosity model when coupled with RANS equations. Similar works should be further done for the machine learning-based full turbulence model considering anisotropy as proposed in this paper.
Furthermore, we will evaluate the proposed network on more complicated, three-dimensional flows. Finally, current input only used vectors; how to include tensor quantities (e.g., strain-rate and rotation-rate) of the mean flow in the input for nonlocal constitutive modeling is an interesting question to explore in future work.

\section*{Acknowledgments}
X.-H. Zhou is supported by the U.S. Air Force under agreement number FA865019-2-2204. The U.S. Government is authorised to reproduce and distribute reprints for Governmental purposes notwithstanding any copyright notation thereon. The computational resources used for this project were provided by the Advanced Research Computing (ARC) of Virginia Tech, which is gratefully acknowledged.

\appendix
\section{Equivariance of the proposed vector-cloud neural network}
\label{app:proof}
We claim that the proposed neural operator has rotational equivariance. That is, given a function \(\mathcal{R} = f(\mathbf{q})\) where $\mathbf{q}$ is the input of vector type and $\mathcal{R}$ is the output of tensor type, we say $f$ is equivariant if the following equation holds for any rotation matrix $\mathsf{Q}$:
\[
\mathsf{Q} \mathcal{R} \mathsf{Q}^\top = f(\mathsf{Q} \mathbf{q})
\qquad \text{or more concisely} \qquad 
\mathcal{R}' = f(\mathbf{q}'),
\]
in which the prime denotes vector or tensor under the rotated frame. Here we demonstrate that a function of the following form satisfies the equivalence condition above:
\begin{equation}
\mathcal{R} = \mathcal{X}^{\top} \cE \mathcal{X} + \gamma \boldsymbol{I} 
\label{eq:equi-form}
\end{equation}
where $\boldsymbol{I}$ is a second order identify tensor, $\mathcal{X}$ are vectors defined in the same coordinate system as $\mathbf{q}$ and $\mathcal{R}$, and $\cE$ is a function with tensor output that is \emph{invariant} with the input vector $\mathbf{q}$, i.e., $\cE(\mathbf{q}) = \cE(\mathbf{q}')$. 
We first note the fact that the identity tensor is equivariant under rotation, i.e., $\mathsf{Q} \boldsymbol{I} \mathsf{Q}^\top = \boldsymbol{I}$. The equivariance can be shown straightforwardly as follows:
\begin{equation} 
\mathsf{Q} \mathcal{R} \mathsf{Q}^\top = 
\mathsf{Q} \mathcal{X}^{\top} \cE \mathcal{X} \mathsf{Q}^\top + \gamma \boldsymbol{I} \\
= \left( \mathsf{Q} \mathcal{X}\right)^{\top} \cE \left(\mathsf{Q} \mathcal{X}\right)^\top
+ \gamma \bm{I}.
\end{equation}
Or more concisely:
\[
\mathcal{R}' = \mathcal{X}' \cE \mathcal{X}'^\top
+ \gamma \bm{I}.
\]
That is, the function form proposed in Eq.~\eqref{eq:equi-form} holds in the rotated frame for any rotation $\mathsf{Q}$.

\section{Interpretation and symmetries of proposed formulation}
\label{app:interp}
The considered formulation of Reynolds stress $\mathcal{R} = \mathcal{X}^{\top} \cE \mathcal{X} + \gamma \boldsymbol{I}$  is apparently similar to the deivatoric--hydrostatic (anisotropic--isotropic) decomposition frequently used in turbulence modeling~\cite{pope00turbulent}:
\begin{equation}
\mathcal{R} \equiv \overline{u_{i} u_{j}}=2 k \left( \bm{b} + \frac{\bm{I}}{3} \right)=2 k\left(\bm{V} \Lambda \mathbf{V}^{\top} + \frac{\bm{I}}{3}\right)
\label{eq:iso}
\end{equation}
where $k$ is the turbulent kinetic energy (half the trace of $\mathcal{R}$), $\bm{b} = \operatorname{dev}(\mathcal{R})/2k$ is the normalized anisotropy, and $\Lambda$ and~$\mathbf{V}$ are eigenvalues and eigenvectors of tensor~$\bm{b}$.
However, it is worth noting that the two forms are fundamentally different. First, the considered decomposition is a mathematical construction to guarantee rotational equivariance. It is not motivated physically. Second, the mathematical structures of the two decompositions are also different. Most notably, the deviatoric tensor $\mathbf{b}$ has trace zero by construction, while \(\mathcal{X}^{\top} \cE \mathcal{X}\) has a nonzero trace. Also note that for flows aligned with the $xy$-plane (normal to $z$-axis), the $xz$-, $yz$-components of the part \(\mathcal{X}^{\top} \cE \mathcal{X}\) are zeros, while both components are nonzero in general for the normalized anisotropic tensor~$\bm{b}$.

It can be seen from the derivation above that a formulation containing only the first part, i.e., $\mathcal{R} = \mathcal{X}^{\top} \cE \mathcal{X}$, would be sufficient to guarantee equivariance of the formulation. Unfortunately, it fails to represent a correct function space for modeling Reynolds stresses in plane mean-strain conditions (i.e., statistically one- or two-dimensional mean flows).
As such, the isotropic part $\gamma \bm{I}$ is added to the formulation to address the functional space mismatch problem, which is further detailed below.

Without loss of generality, consider a turbulent flow with the two-dimensional mean flow in the $xy$-plane (i.e., the $z$-direction is statistically homogeneous and thus has zero strain-rate). The $z$-component of the relative coordinates $\mathcal{X}$ is zero. In this case simple algebra shows that the tensor $\mathcal{X}^{\top} \cE \mathcal{X}$ has the following form:
\[
\begin{bmatrix}
\rstt_{xx} & \rstt_{xy} & 0 \\
 & \rstt_{yy} & 0 \\
\text{symm} &  & 0
\end{bmatrix}
\]
It is noteworthy that by construction the formulation correctly ensures that (1) the output  is a symmetric tensor and (2) that the shear components $\rstt_{xz}$ and $\rstt_{yz}$ terms are zero in such two-dimensional mean flows. However, the normal stress component $\rstt_{zz}$ is also forced to zero, while in general this term is nonzero even in statistically two-dimensional flows (or even in the simplest isotropic homogeneous flows). For example, it has been shown by experiments that the ratios of normal stress components (i.e., turbulence intensities) in the streamwise ($x$-), wall-normal ($y$-), and spanwise ($z$-) directions are $\rstt_{xx} : \rstt_{yy} : \rstt_{zz} \approx 1 : 0.4 : 0.6$ in the log-law region of a zero-pressure gradient flat-late boundary layer~\cite{klebanoff1955characteristics}.

One would be tempted to remedy the inadvertent singularity by generating the training data in a general plane (i.e., by rotating the coordinate frame so that the flow is not aligned with the $xy$-plane). However, this is not a valid workaround. Since the formulation is rotational equivariant by construction, no matter in which coordinate system the training is performed, the trained model will still predict $\rstt_{zz}=0$ if the flow is aligned with the $xy$-plane, which violates the physics discussed above.

There is also another perspective to understand the inadequacy of the term $\mathcal{X}^{\top} \cE \mathcal{X}$. When the $z$-component of the relative coordinate $\mathcal{X}$ is always zero, $\mathcal{X}$'s row space is at most two-dimensional. Note that the row space of $\mathcal{X}^{\top} \cE \mathcal{X}$ is no larger than that of $\mathcal{X}$, i.e., it has a rank of no more than two. In contrast, the physical Reynolds stress has a full rank (i.e., of three) in general even for statistically one- or two-dimensional mean flows. Hence, the formulation $\mathcal{R} = \mathcal{X}^{\top} \cE \mathcal{X}$ cannot adequately represent Reynolds stresses in one- or two-dimensional mean flows.

In view of such limitations of $\mathcal{X}^{\top} \cE \mathcal{X}$, we propose to add a term that allows the $\rstt_{zz}$ component to be nonzero, while preserving the equivariance of the first part ($\mathcal{X}^{\top} \cE \mathcal{X}$) as well as the existing symmetries of the output tensor (i.e., $\rstt_{yz} = \rstt_{yz} = 0$ and $\rstt = \rstt^\top$).  The term $\gamma \bm{I}$ satisfies all these requirements. Note that here $\gamma$ is a trainable scalar parameter as opposed to $2k/3$ in the anisotropic decomposition~Eq.~\eqref{eq:iso}. Finally, because the formulation is equivariant, such a construction is a valid remedy regardless of the direction of the plane strain.

\section{Network architectures and training parameters}
\label{app:network-architecture}
Detailed architecture of VCNN-e, consisting of an embedding network and a fitting network, is provided in Table~\ref{tab:nn-details}. The embedding network operates identically on the scalar features $\bm{c}\in \bbR^{l'}$ associated with each point in the cloud and outputs a row of $m$ elements in matrix~$\cG$. The fitting network maps the invariant feature matrix $\mathcal{D}$ to a diagonal matrix $\mathcal{E}$ and a single element $\gamma$.

The neural networks are trained using the Adam optimizer~\cite{kingma2015adam}. Both training processes take 5000 epochs, with batch sizes of 512 for synthetic data and 256 for DNS data, respectively. The learning rate is set to be 0.001 at the begining of the training process.

\begin{table}[htbp]
\caption{Detailed architectures of the embedding network and the fitting network. The architecture specifies the number of neurons in each layer in sequence, from the input layer to the hidden layers (if any) and the output layer. The numbers of neurons in the input and output layers are highlighted in bold. Note that the number of input neurons in the embedding network for the DNS scenario changes to $l^{\prime}=8$ with an additional scalar feature of mean pressure.
}
\centering
\begin{tabular}{p{4.5cm} p{4.5cm} p{4.5cm}}
\toprule[1pt]
 & Embedding network  & Fitting network ($\mathcal{D} \mapsto \mathcal{E},\gamma$)\\
\midrule
No. of input neurons & $l' = 7 $ & $m\times m'= 256 \, (\cD \in \bbR^{64\times 4})$\\
No. of hidden layers & 3 & 2 \\
Architecture & 
(\textbf{7}, 32, 64, 64, \textbf{64}) 
& 
(\textbf{256}, 64, 64, \textbf{65})
\\
No. of output neurons & $m=64$ & 65 ($\mathcal{E} \in \bbR^{64\times64}$, $\gamma \in \bbR$) \\
Activation functions & ReLU, Linear (last layer) & ReLU, Linear (last layer) \\
No. of trainable parameters 
& 10688 & 24833
\\
\bottomrule[1pt]
\end{tabular}
\label{tab:nn-details}
\end{table}

\section{Progressive training of neural networks}
\label{app:training-method-compare}
\begin{figure}[htb]
\centering
\includegraphics[width=0.9\linewidth]{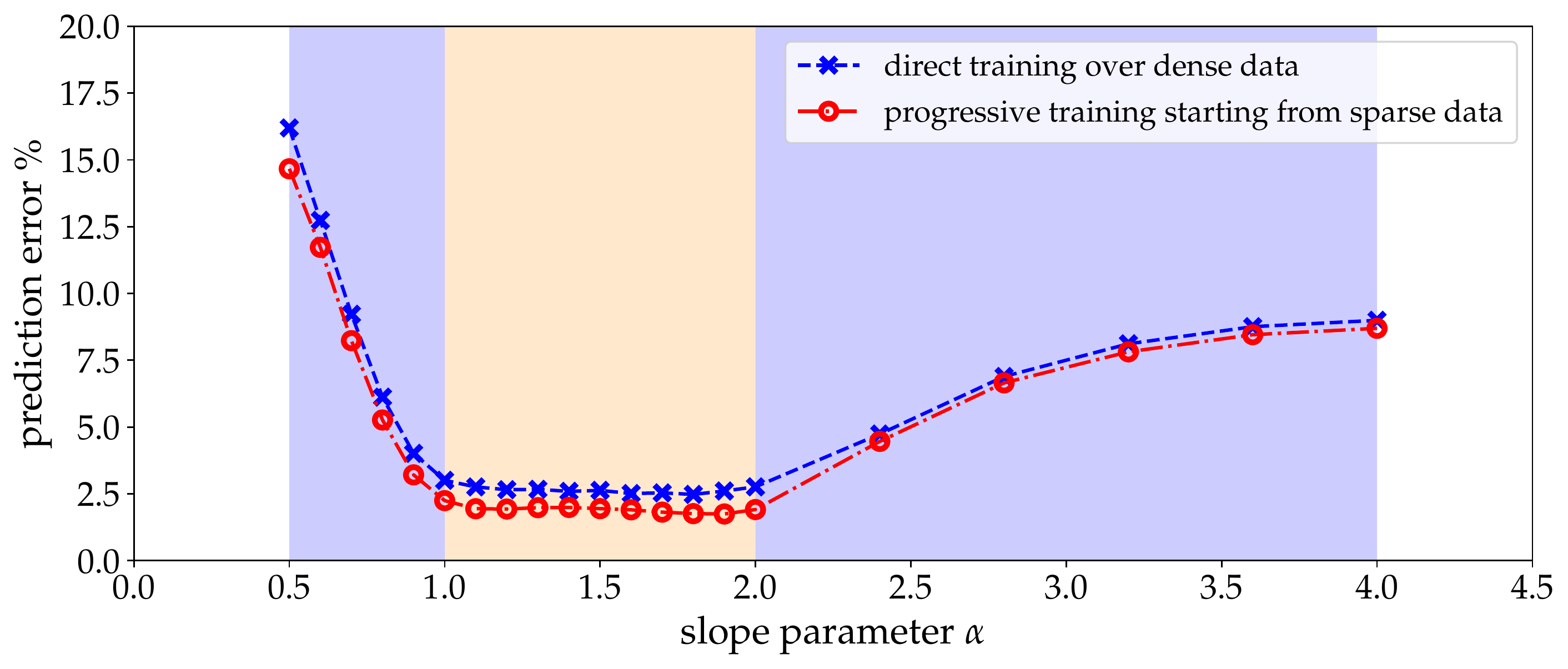}
  \caption{
  Comparison of direct and progressive training methods in terms of prediction performance for turbulence kinetic energy $k$ at various slope parameters $\alpha$ between 0.5 and 4. The blue dashed and red dash-dotted lines represent the prediction errors of the trained models by direct and progressive training methods, respectively. The testings on 21 flows are based on all the data points within the cloud, and thus the comparison is fair. 
  \label{fig:prediction-error-compare}
  }
\end{figure}

In this work, we have explored two training methods: (1) direct training that always uses dense data, with each cloud having $n=300$ points, and (2) progressive training first with sparse data ($n=25$ points in the cloud) and then dense data. 
In the first method, we directly train the model with 5000 epochs. In the second method, we first train the model using the sparse data for 1000 epochs and then using the dense data for 4000 epochs.
Under the same number of epochs, the second strategy reduces training time by approximately 18\% compared to the first method, with an average of 11.12 and 13.56 seconds per epoch, respectively.
The prediction performances of the trained models under two different methods are compared over 21 different configurations, which is shown in Fig.~\ref{fig:prediction-error-compare}. 
We can see that, with less training time, the progressive method performs better in all testing cases. This is because the first stage of training using sparse data (i.e., a subset of full training data) accelerates the convergence of the neural operator. Note that for progressive training, the number of training epochs with sparse data is not fine-tuned. We select 1000 epochs as an example to demonstrate how progressive training improves the training efficiency. More settings can be investigated to optimize the progressive training.

\section{Predicted Reynolds stress anisotropy invariant $A_2$ in extrapolated flows}
\label{app:extra-kA2}

\begin{figure*}[!htb]
\centering
{\includegraphics[width=0.9\linewidth]{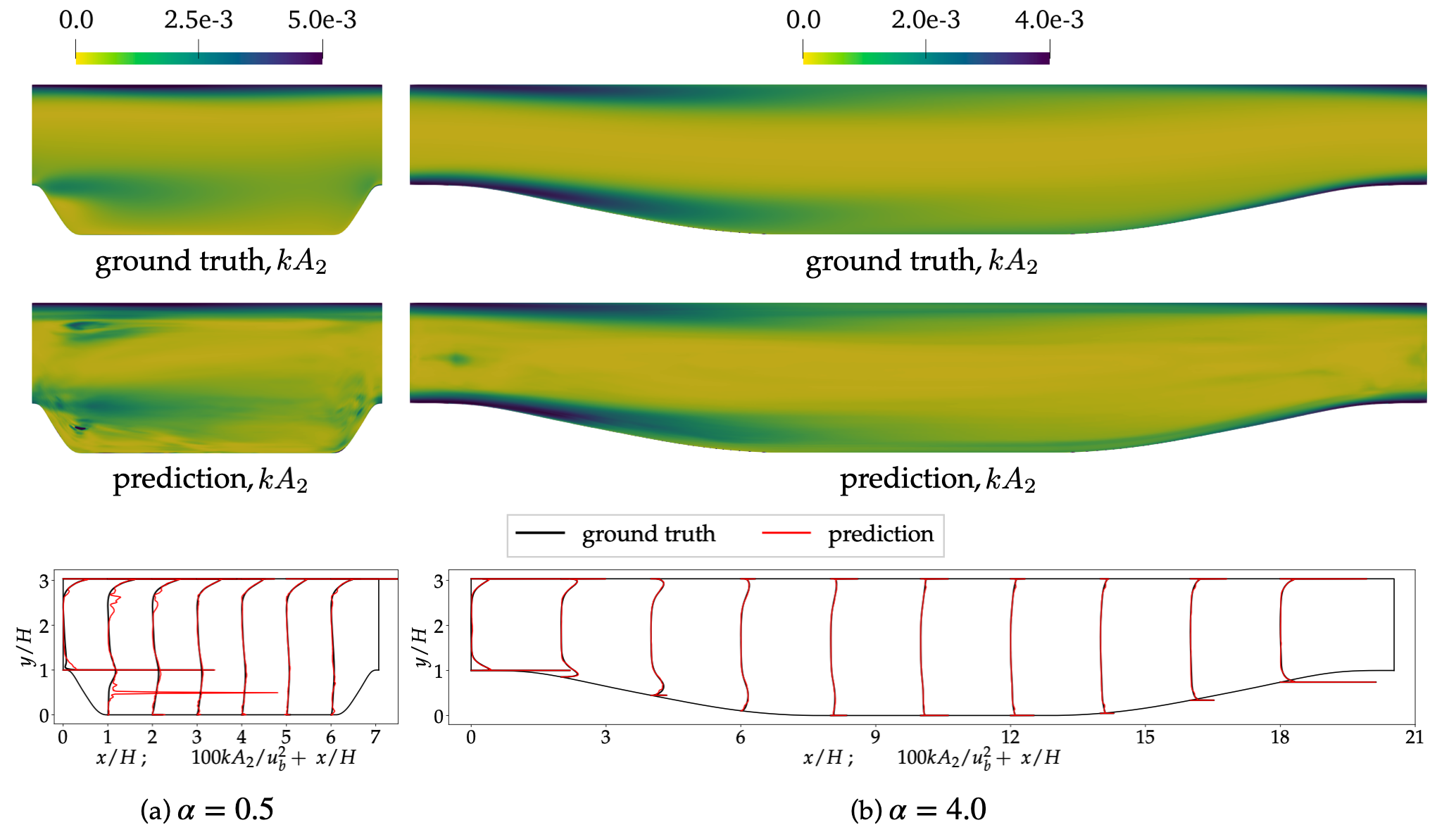}}
  \caption{Comparison of the ground truths of $kA_2$, the \textbf{Reynolds stress anisotropy invariant} scaled by the turbulent kinetic energy (top row) and the corresponding predictions based on the predicted Reynolds stress with the trained neural network (middle row), along with the scaled $A_2$ profiles at seven and ten cross-sections (bottom row) for two extreme \textbf{extrapolated} configurations with slope parameters $\alpha=0.5$ (left panels) and $\alpha =4.0$ (right panels).
  }
  \label{fig:A2-extrapolation-compare}
\end{figure*}

Predictions of anisotropy invariant $A_2$ in two extreme extrapolated flows with $\alpha=0.5$ and $\alpha=4$ are illustrated in Fig.~\ref{fig:A2-extrapolation-compare} as they show the same general trend as those in Section~\ref{sec:res-tau} while providing some additional evidence on the performance of the proposed method. In the flow with $\alpha=0.5$, the location with distinct inconsistency is still found at the leeside of the hill, which is shown more clearly with a 50 times magnification at cross-sections. The fast contraction and separation of the flow near hill results in rapid changes in strain and rotation rates; this extrapolation error is further amplified by the square operation in the definition of $A_2$. In the flow with $\alpha=4$, the predicted $k A_2$ is more accurate as there is virtually no flow contraction and separation due to the gentle slope.

\section{Predictions of anisotropy states}
\label{app:anisotropy}

\begin{figure}[t]
\centering
\includegraphics[width=0.88\linewidth]{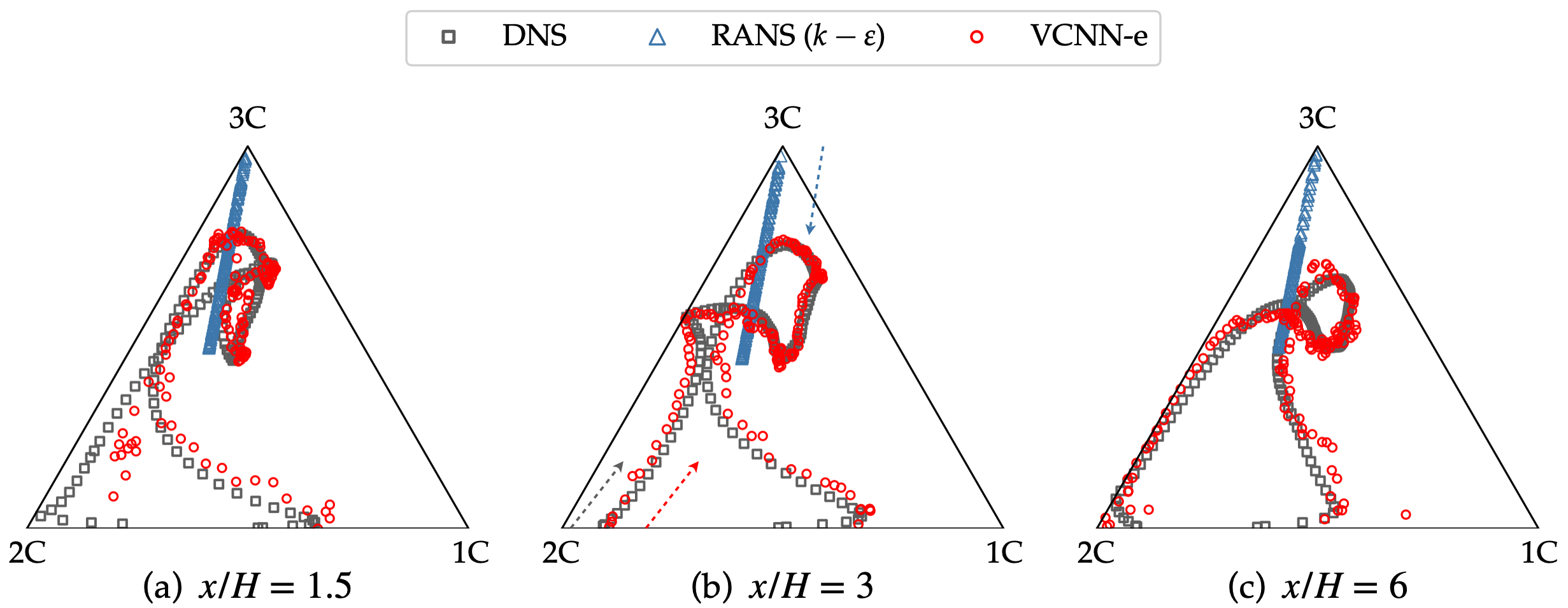}
\caption{Predicted Reynolds stress anisotropy for testing configuration with slope parameter $\alpha=1.0$. Prediction from the nonlocal model (VCNN-e) is compared with the ground truth (DNS) and $k-\varepsilon$ RANS result at three different cross-sections. The Reynolds stress anisotropy componentality is represented by its position in the Barycentric: one-component state (1C), two-component isotropic state (2C), and three-component isotropic state (3C), as well as a combination thereof (interior). The dashed arrows indicate the direction from the bottom wall to the top wall.}
\label{fig:DNS-triangle}
\end{figure}

We examine the anisotropy states of the predicted Reynolds stress in Section~\ref{sec:results_dns}, which exhibit a high degree of similarity with the original DNS data as well. The anisotropy state is represented by the position within the Barycentric triangle shown in Fig.~\ref{fig:DNS-triangle}. The three vertices, 1C, 2C, and 3C, of the Barycentric triangle represent, in order, the one-component state, the two-component isotropic state, and the three-component isotropic state. The interior position represents a combination thereof. The coordinates of a location in the Barycentric triangle are determined by the eigenvalues of the normalized Reynolds stress anisotropy $\bm{b}$. More details on the calculation of coordinates in the Barycentric triangle can be found in the paper~\cite{xiao2020flows}.

As illustrated in Fig.~\ref{fig:DNS-triangle}, the positions of the predicted anisotropy states within the Barycentric triangle are comparable to those of the DNS data along three cross-sections ($x/H = 1.5, 3, 6$). Such consistency demonstrates that the VCNN-e is capable of providing a reasonable anisotropy prediction for an unseen configuration despite the fact that the anisotropy is not explicitly included in the loss function. At cross-section $x/H=1.5$, the near-wall region close to the hill displays the most apparent discrepancy. This is because the region lies within the recirculation zone, where the different hill slope causes the mean flow field to differ from that in the training cases. Moreover, the positions of predicted turbulence states from VCNN-e show better componentality than the RANS result. The positions of near-wall turbulence states from the standard $k-\varepsilon$ RANS model are observed near the three-component isotropic state (vertex 3C), whereas both predicted and actual positions are indicated to be close to the two-component state (bottom edge).

\clearpage

\bibliographystyle{elsarticle-num}

\begin{thebibliography}{10}
\expandafter\ifx\csname url\endcsname\relax
  \def\url#1{\texttt{#1}}\fi
\expandafter\ifx\csname urlprefix\endcsname\relax\def\urlprefix{URL }\fi
\expandafter\ifx\csname href\endcsname\relax
  \def\href#1#2{#2} \def\path#1{#1}\fi

\bibitem{spalart92one}
P.~R. Spalart, S.~R. Allmaras, A one equation turbulence model for aerodynamic
  flows., AIAA Journal 94 (1992).

\bibitem{launder74application}
B.~Launder, B.~Sharma, Application of the energy-dissipation model of
  turbulence to the calculation of flow near a spinning disc, Letters in Heat
  and Mass Transfer 1~(2) (1974) 131--137.

\bibitem{launder1975progress}
B.~E. Launder, G.~J. Reece, W.~Rodi, Progress in the development of a
  {Reynolds-stress} turbulence closure, Journal of Fluid Mechanics 68~(3)
  (1975) 537--566.

\bibitem{coleman1967thermodynamics}
B.~D. Coleman, M.~E. Gurtin, Thermodynamics with internal state variables, The
  Journal of Chemical Physics 47~(2) (1967) 597--613.

\bibitem{han2019uniformly}
J.~Han, C.~Ma, Z.~Ma, W.~E, Uniformly accurate machine learning-based
  hydrodynamic models for kinetic equations, Proceedings of the National
  Academy of Sciences 116~(44) (2019) 21983--21991.

\bibitem{lei2020machine}
H.~Lei, L.~Wu, W.~E, Machine-learning-based non-newtonian fluid model with
  molecular fidelity, Physical Review E 102~(4) (2020) 043309.

\bibitem{han2021machine}
W.~E, J.~Han, L.~Zhang, Machine-learning-assisted modeling, Physics Today
  74~(7) (2021) 36--41.

\bibitem{ling16reynolds}
J.~Ling, A.~Kurzawski, J.~Templeton, Reynolds averaged turbulence modelling
  using deep neural networks with embedded invariance, Journal of Fluid
  Mechanics 807 (2016) 155--166.

\bibitem{wang17physics-informed}
J.-X. Wang, J.-L. Wu, H.~Xiao, Physics-informed machine learning approach for
  reconstructing {Reynolds} stress modeling discrepancies based on {DNS} data,
  Physical Review Fluids 2~(3) (2017) 034603.

\bibitem{wu2018physics-informed}
J.-L. Wu, H.~Xiao, E.~G. Paterson, Physics-informed machine learning approach
  for augmenting turbulence models: A comprehensive framework, Physical Review
  Fluids 3 (2018) 074602.

\bibitem{schmelzer2020discovery}
M.~Schmelzer, R.~P. Dwight, P.~Cinnella, Discovery of algebraic reynolds-stress
  models using sparse symbolic regression, Flow, Turbulence and Combustion
  104~(2) (2020) 579--603.

\bibitem{bock2019review}
F.~E. Bock, R.~C. Aydin, C.~J. Cyron, N.~Huber, S.~R. Kalidindi, B.~Klusemann,
  A review of the application of machine learning and data mining approaches in
  continuum materials mechanics, Frontiers in Materials 6 (2019) 110.

\bibitem{huang2020learning}
D.~Z. Huang, K.~Xu, C.~Farhat, E.~Darve, Learning constitutive relations from
  indirect observations using deep neural networks, Journal of Computational
  Physics (2020) 109491.

\bibitem{xu2021learning}
K.~Xu, D.~Z. Huang, E.~Darve, Learning constitutive relations using symmetric
  positive definite neural networks, Journal of Computational Physics 428
  (2021) 110072.

\bibitem{masi2021thermodynamics}
F.~Masi, I.~Stefanou, P.~Vannucci, V.~Maffi-Berthier, Thermodynamics-based
  artificial neural networks for constitutive modeling, Journal of the
  Mechanics and Physics of Solids 147 (2021) 104277.

\bibitem{scoggins2021machine}
J.~B. Scoggins, J.~Han, M.~Massot, Machine learning moment closures for
  accurate and efficient simulation of polydisperse evaporating sprays, in:
  AIAA Scitech 2021 Forum, 2021, p. 1786.

\bibitem{wu2019representation}
J.-L. Wu, R.~Sun, S.~Laizet, H.~Xiao, Representation of stress tensor
  perturbations with application in machine-learning-assisted turbulence
  modeling, Computer Methods in Applied Mechanics and Engineering 346 (2019)
  707--726.

\bibitem{zafar2020convolutional}
M.~I. Zafar, H.~Xiao, M.~M. Choudhari, F.~Li, C.-L. Chang, P.~Paredes,
  B.~Venkatachari, Convolutional neural network for transition modeling based
  on linear stability theory, Physical Review Fluids 5 (2020) 113903.

\bibitem{taghizadeh2020turbulence}
S.~Taghizadeh, F.~D. Witherden, S.~S. Girimaji, Turbulence closure modeling
  with data-driven techniques: physical compatibility and consistency
  considerations, New Journal of Physics 22~(9) (2020) 093023.

\bibitem{wang2021incorporating}
R.~Wang, R.~Walters, R.~Yu, Incorporating symmetry into deep dynamics models
  for improved generalization, in: International Conference on Learning
  Representations, 2021.

\bibitem{pope1975more}
S.~B. Pope, A more general effective-viscosity hypothesis, Journal of Fluid
  Mechanics 72~(2) (1975) 331--340.

\bibitem{gatski1993on}
T.~Gatski, C.~Speziale, On explicit algebraic stress models for complex
  turbulent flows, Journal of Fluid Mechanics 254 (1993) 59--79.

\bibitem{speziale1991modelling}
C.~G. Speziale, S.~Sarkar, T.~B. Gatski, Modelling the pressure--strain
  correlation of turbulence: an invariant dynamical systems approach, Journal
  of Fluid Mechanics 227 (1991) 245--272.

\bibitem{gatski1996simulation}
T.~B. Gatski, M.~Y. Hussaini, J.~L. Lumley, Simulation and modeling of
  turbulent flows, Oxford University Press, 1996.

\bibitem{gao2020roteqnet}
L.~Gao, Y.~Du, H.~Li, G.~Lin, {RotEqNet}: rotation-equivariant network for
  fluid systems with symmetric high-order tensors, Journal of Computational
  Physics 461 (2022) 111205.

\bibitem{long2018pde}
Z.~Long, Y.~Lu, X.~Ma, B.~Dong, {PDE}-{N}et: Learning {PDE}s from data, in:
  International Conference on Machine Learning, PMLR, 2018, pp. 3208--3216.

\bibitem{sun2020surrogate}
L.~Sun, H.~Gao, S.~Pan, J.-X. Wang, Surrogate modeling for fluid flows based on
  physics-constrained deep learning without simulation data, Computer Methods
  in Applied Mechanics and Engineering 361 (2020) 112732.

\bibitem{kim19deep}
B.~Kim, V.~C.~Azevedo, N.~Thuerey, T.~Kim, M.~Gross, B.~Solenthaler,
  {DeepFluids}: A generative network for parameterized fluid simulations,
  Computer Graphics Forum (Proc. Eurographics) 38~(2) (2019).

\bibitem{lu2021learning}
L.~Lu, P.~Jin, G.~Pang, Z.~Zhang, G.~E. Karniadakis, Learning nonlinear
  operators via {DeepONet} based on the universal approximation theorem of
  operators, Nature Machine Intelligence 3~(3) (2021) 218--229.

\bibitem{ma2020machine}
C.~Ma, B.~Zhu, X.-Q. Xu, W.~Wang, Machine learning surrogate models for
  {L}andau fluid closure, Physics of Plasmas 27~(4) (2020) 042502.

\bibitem{ribeiro2020deepcfd}
M.~D. Ribeiro, A.~Rehman, S.~Ahmed, A.~Dengel, {DeepCFD}: Efficient
  steady-state laminar flow approximation with deep convolutional neural
  networks, arXiv preprint arXiv:2004.08826 (2020).

\bibitem{li2020neural}
Z.~Li, N.~Kovachki, K.~Azizzadenesheli, B.~Liu, K.~Bhattacharya, A.~Stuart,
  A.~Anandkumar, Neural operator: Graph kernel network for partial differential
  equations, arXiv preprint arXiv:2003.03485 (2020).

\bibitem{li2021fourier}
Z.~Li, N.~B. Kovachki, K.~Azizzadenesheli, B.~liu, K.~Bhattacharya, A.~Stuart,
  A.~Anandkumar, Fourier neural operator for parametric partial differential
  equations, in: International Conference on Learning Representations, 2021.

\bibitem{han2018deep}
J.~Han, L.~Zhang, R.~Car, W.~E, Deep {P}otential: a general representation of a
  many-body potential energy surface, Communications in Computational Physics
  23~(3) (2018) 629--639.

\bibitem{zhang2018end}
L.~Zhang, J.~Han, H.~Wang, W.~Saidi, R.~Car, W.~E, End-to-end symmetry
  preserving inter-atomic potential energy model for finite and extended
  systems, in: Advances in neural information processing systems, 2018, pp.
  4436--4446.

\bibitem{zhou2022frame}
X.-H. Zhou, J.~Han, H.~Xiao, Frame-independent vector-cloud neural network for
  nonlocal constitutive modeling on arbitrary grids, Computer Methods in
  Applied Mechanics and Engineering 388 (2022) 114211.

\bibitem{sommers2020raman}
G.~M. Sommers, M.~F.~C. Andrade, L.~Zhang, H.~Wang, R.~Car, Raman spectrum and
  polarizability of liquid water from deep neural networks, Physical Chemistry
  Chemical Physics 22~(19) (2020) 10592--10602.

\bibitem{xiao2020flows}
H.~Xiao, J.-L. Wu, S.~Laizet, L.~Duan, Flows over periodic hills of
  parameterized geometries: A dataset for data-driven turbulence modeling from
  direct simulations, Computers \& Fluids 200 (2020) 104431.

\bibitem{launder75progress}
B.~Launder, G.~J. Reece, W.~Rodi, Progress in the development of a
  {Reynolds}-stress turbulence closure, Journal of Fluid Mechanics 68~(03)
  (1975) 537--566.

\bibitem{pope00turbulent}
S.~B. Pope, Turbulent Flows, Cambridge University Press, Cambridge, 2000.

\bibitem{opencfd21openfoam}
{The OpenFOAM Foundation},
  \href{https://cfd.direct/openfoam/user-guide}{{OpenFOAM} User Guide} (2020).
\newline\urlprefix\url{https://cfd.direct/openfoam/user-guide}

\bibitem{issa86solution}
R.~I. Issa, Solution of the implicitly discretised fluid flow equations by
  operator-splitting, Journal of Computational Physics 62 (1986) 40--65.

\bibitem{zhou2021learning}
X.-H. Zhou, J.~Han, H.~Xiao, Learning nonlocal constitutive models with neural
  networks, Computer Methods in Applied Mechanics and Engineering 384 (2021)
  113927.

\bibitem{paszke2019pytorch}
A.~Paszke, S.~Gross, F.~Massa, A.~Lerer, J.~Bradbury, G.~Chanan, T.~Killeen,
  Z.~Lin, N.~Gimelshein, L.~Antiga, et~al., Pytorch: An imperative style,
  high-performance deep learning library, in: Advances in neural information
  processing systems, 2019, pp. 8026--8037.

\bibitem{zhou2022vcnne-git}
X.-H. Zhou, J.~Han, H.~Xiao, Vector-cloud neural network with equivariance
  ({VCNN-e}) for tensors,
  \url{https://github.com/xuhuizhou-vt/VCNN-nonlocal-constitutive-model/tree/master/tensor-transport}.

\bibitem{lumley1977return}
J.~L. Lumley, G.~R. Newman, The return to isotropy of homogeneous turbulence,
  Journal of Fluid Mechanics 82~(1) (1977) 161--178.

\bibitem{xiaoperiodichill-git}
H.~Xiao, S.~Laizet, Periodic hill database,
  \url{https://github.com/xiaoh/para-database-for-PIML/tree/master/pehill_new_DNS_database}.

\bibitem{laizet2011incompact3d}
S.~Laizet, N.~Li, Incompact3d: A powerful tool to tackle turbulence problems
  with up to {O}($10^5$) computational cores, International Journal for
  Numerical Methods in Fluids 67~(11) (2011) 1735--1757.

\bibitem{launder1974application}
B.~Launder, B.~Sharma, Application of the energy-dissipation model of
  turbulence to the calculation of flow near a spinning disc, Letters in heat
  and mass transfer 1~(2) (1974) 131--137.

\bibitem{zafar2021frame}
M.~I. Zafar, J.~Han, X.-H. Zhou, H.~Xiao, Frame invariance and scalability of
  neural operators for partial differential equations, Communications in
  Computational Physics 32~(2) (2022) 336--363.

\bibitem{xu2022pde}
R.~Xu, X.-H. Zhou, J.~Han, R.~P. Dwight, H.~Xiao, A {PDE}-free, neural
  network-based eddy viscosity model coupled with {RANS} equations,
  International Journal of Heat and Fluid Flow 98 (2022) 109051.

\bibitem{klebanoff1955characteristics}
P.~S. Klebanoff, Characteristics of turbulence in a boundary layer with zero
  pressure gradient, Tech. rep., National Bureau of Standards, Gaithersburg, MD
  (1955).

\bibitem{kingma2015adam}
D.~Kingma, J.~Ba, Adam: a method for stochastic optimization, in: Proceedings
  of the International Conference on Learning Representations, 2015.

\end{thebibliography}

 \clearpage
 \thenomenclature
 \nomgroup{A}
 \item [{$\nabla$}]\begingroup spatial derivatives
 \item [{$\hat{\boxed{}}$}]\begingroup prediction
 \item [{${\boxed{}}^*$}]\begingroup ground truth
 \item [{${\boxed{}}^\star$}]\begingroup submatrix
 \item [{$|{\cdot}|$}]\begingroup absolute value, vector magnitude
 \item [{$\|{\cdot}\|$}]\begingroup tensor magnitude
 \item [{$\region{\cdot}$}]\begingroup variables in a nonlocal region
 \item [{$\boxed{}^\top$}]\begingroup matrix transpose
 \item [{$\operatorname{dev}(\cdot)$}]\begingroup deviatoric component of a tensor

 \nomgroup{B}
 \item [{$A_2$}]\begingroup Reynolds stress anisotropy invariant
 \item [{$b$}]\begingroup boundary cell indicator (input feature)
 \item [{$\bm{b}$}]\begingroup normalized Reynolds stress anisotropy
 \item [{$C_\mu$, $C_D$, $C_1$, $C_2$}]\begingroup \quad model coefficients
 \item [{$D$}]\begingroup elements in invariant feature matrix $\mathcal{D}$
 \item [{$\mathcal{D}$}]\begingroup invariant feature matrix
 \item [{$e$}]\begingroup diagonal elements in invariant diagonal matrix
 \item [{$E$}]\begingroup dissipation term
 \item [{$\mathcal{E}$}]\begingroup invariant diagonal matrix
 \item [{$\mathcal{G}$}]\begingroup embedding matrix
 \item [{$H$}]\begingroup crest height
 \item [{$\boldsymbol{I}$}]\begingroup identity tensor
 \item [{$k$}]\begingroup turbulent kinetic energy
 \item [{$\ell_1$, $\ell_2$}]\begingroup length of semi-major and semi-minor axes of the nonlocal region
 \item [{$\ell_m$}]\begingroup mixing length
 \item [{$\ell'$}]\begingroup number of input neurons of the embedding netowrk
 \item [{$L_x$}]\begingroup computational domain length
 \item [{$L_y$}]\begingroup computational domain height
 \item [{$\mathcal{L}$}]\begingroup permutational invariant matrix
 \item [{$m$}]\begingroup number of embedding functions
 \item [{$m'$}]\begingroup number of a subset of embedding functions
 \item [{$n$}]\begingroup stencil size (number of sampled points in the cloud)
 \item [{$p$}]\begingroup pressure
 \item [{$\mathcal{P}$}]\begingroup production term
 \item [{$\mathbf{q}$}]\begingroup feature vector
 \item [{$\mathcal{Q}$}]\begingroup input matrix
 \item [{$\mathsf{Q}$}]\begingroup rotation matrix
 \item [{$\mathcal{R}$}]\begingroup Reynolds stress tensor
 \item [{$Re$}]\begingroup Reynolds number
 \item [{$r$}]\begingroup inverse of relative distance to the cloud center (input feature)
 \item [{$r'$}]\begingroup alignment between the convection and the relative position of non-local points (input feature)
 \item [{$\bm{S}$}]\begingroup strain rate tensor
 \item [{$s$}]\begingroup magnitude of strain rate
 \item [{$t$}]\begingroup temporal coordinate
 \item [{$u$}]\begingroup mean velocity in $x$ direction
 \item [{$v$}]\begingroup mean velocity in $y$ direction
 \item [{$\mathbf{u}$}]\begingroup mean velocity field
 \item [{$\mathsf{u}$}]\begingroup velocity magnitude (input feature)
 \item [{$u_b$}]\begingroup bulk velocity magnitude
 \item [{$\bm{V}$}]\begingroup eigenvectors of tensor $\bm{b}$
 \item [{$w$}]\begingroup hill width
 \item [{$\bm{x}$}]\begingroup spatial coordinate
 \item [{$\bm{x}_0$}]\begingroup spatial coordinate of cloud center
 \item [{$\bm{x}'$}]\begingroup relative spatial coordinate to the cloud center
 \item [{$\tilde{\mathcal{X}}$}]\begingroup embedded coordinates
 \nomgroup{C}
 \item [{$\alpha$}]\begingroup slope parameter
 \item [{$\gamma$}]\begingroup extra output of fitting network
 \item [{$\delta_{ij}$}]\begingroup Kronecker delta
 \item [{$\delta^*$}]\begingroup boundary layer thickness
 \item [{$\epsilon$}]\begingroup error tolerance
 \item [{$\varepsilon$}]\begingroup dissipation rate scalar
 \item [{$\zeta$}]\begingroup dissipation coefficient
 \item [{$\eta$}]\begingroup wall distance (input feature)
 \item [{$\theta$}]\begingroup cell volume (input feature)
 \item [{$\kappa$}]\begingroup von Karman constant
 \item [{$\Lambda$}]\begingroup eigenvalues of tensor $\bm{b}$
 \item [{$\nu$}]\begingroup kinematic viscosity
 \item [{$\nu_t$}]\begingroup turbulent eddy viscosity
 \item [{$\mu$}]\begingroup dynamic viscosity
 \item [{$\rho$}]\begingroup density
 \item [{$\tau$}]\begingroup turbulence time scale
 \item [{$\Phi$}]\begingroup pressure--strain-rate term
 \item [{$\phi$}]\begingroup basis function of embedding neural network
 \item [{$\bm{\Omega}$}]\begingroup rotation rate tensor

\end{document}